\documentclass[11pt,a4paper]{article}
\usepackage{jcappub}

\usepackage{amsmath,amssymb,bm,float,mathrsfs,subfigure,tabularx}
\usepackage{dcolumn}
\usepackage{color}
  \newcommand{\Blue}[1]{\textcolor{blue}{#1}}
  \newcommand{\beq}{\begin{equation}}
  \newcommand{\eeq}{\end{equation}}
  \newcommand{\al}[1]{\begin{align} #1 \end{align}}
  \newcommand{\bi}{\begin{itemize}}
  \newcommand{\ei}{\end{itemize}}
  \newcommand{\bc}{\begin{center}}
  \newcommand{\ec}{\end{center}}
  \newcommand{\rom}[1]{_{\mathrm{#1}}}
  \def\dd{\mathrm{d}}

  \def\rmm{\mathrm{m}}
  
  \def\rms{\mathrm{s}}

  \def\rmD{\mathrm{D}}

  \def\rmS{\mathrm{S}}

  \def\mcA{\mathcal{A}}

  \def\mcO{\mathcal{O}}

  \newcommand{\ave}[1]{\left\langle #1 \right\rangle}

  \def\3Dint#1{\int\frac{\dd^{3}{#1 }}{(2\pi )^3}}
  \def\cCMB{\chi_{\rm CMB}}

\title{CMB ISW-lensing bispectrum from cosmic strings}

\author[a]{Daisuke Yamauchi}
\author[b]{Yuuiti Sendouda}
\author[c]{Keitaro Takahashi}

\affiliation[a]{%
Research Center for the Early Universe, Graduate School of Science, 
The University of Tokyo, Bunkyo-ku, Tokyo 113-0033, Japan
}%
\affiliation[b]{%
Graduate School of Science and Technology, Hirosaki University, Hirosaki, Aomori 036-8561, Japan
}%
\affiliation[c]{%
Faculty of Science, Kumamoto University, 2-39-1 Kurokami, Kumamoto 860-8555, Japan
}%

\emailAdd{yamauchi@resceu.s.u-tokyo.ac.jp}

\abstract{
We study the effect of weak lensing by cosmic (super-)strings on the higher-order statistics of the cosmic microwave background (CMB).
A cosmic string segment is expected to cause weak lensing as well as an integrated Sachs-Wolfe (ISW) effect, the so-called Gott-Kaiser-Stebbins (GKS) effect, to the CMB temperature fluctuation, which are thus naturally cross-correlated.
We point out that, in the presence of such a correlation, yet another kind of the post-recombination CMB temperature bispectra, the ISW-lensing bispectra, will arise in the form of products of the auto- and cross-power spectra.
We first present an analytic method to calculate the autocorrelation of the temperature fluctuations induced by the strings, and the cross-correlation between the temperature fluctuation and the lensing potential both due to the string network.
In our formulation, the evolution of the string network is assumed to be characterized by the simple analytic model, 
the velocity-dependent one scale model, and the intercommutation probability is properly incorporated in order 
to characterize the possible superstringy nature.
Furthermore, the obtained power spectra are dominated by the Poisson-distributed string segments,
whose correlations are assumed to satisfy the simple relations.
We then estimate the signal-to-noise ratios of the string-induced ISW-lensing bispectra and discuss the detectability of such CMB signals from the cosmic string network.
It is found that in the case of the smaller string tension, $G\mu\ll 10^{-7}$\,, 
the ISW-lensing bispectrum induced by a cosmic string network can constrain the string-model parameters even more tightly than the purely GKS-induced bispectrum in the ongoing and future CMB observations on small scales.
}

\begin{document}

\maketitle

%%%%%%%%%%%%%%%%%%%%%%%%%%%%%%%%%%%%%%%%%%%%%%%%%%%%%%%%%%%%%%%%%%%%%%%%%
%=======================================================================%
\section{Introduction}
\label{sec:Introduction}
%=======================================================================%
%%%%%%%%%%%%%%%%%%%%%%%%%%%%%%%%%%%%%%%%%%%%%%%%%%%%%%%%%%%%%%%%%%%%%%%%%

Topological defects, appearing as solutions to the field equation in various models of particle physics, are expected to have formed during phase transitions in the early universe through spontaneous symmetry breakings~\cite{Kibble:1976sj,Zeldovich:1980gh,Vilenkin:1981iu} (see \cite{Vilenkin-Shellard} for a review).
It has been shown that cosmic strings generally appear at the end of inflation within a various variety of supersymmetric grand unified theories~\cite{Jeannerot:2003qv}.

In the late-time universe, intercommutation of cosmic strings serves as an essential mechanism of energy dissipation which keeps the total energy of strings within the expanding Hubble volume from growing.
Early studies on this subject~\cite{Vilenkin:1981iu,Kibble:1984hp} employed analytic methods and suggested formation of a stable structure with constant energy density, so-called \textit{scaling string network}.
Afterwards, numerical simulations of dynamical formation of a string network in the expanding universe have been performed for the Nambu-Goto strings~\cite{Albrecht:1984xv,Bennett:1989yp,Albrecht:1989mk,Allen:1990tv,Vincent:1996rb,Martins:2005es,Olum:2006ix,Ringeval:2005kr,Fraisse:2007nu,BlancoPillado:2011dq} and Abelian-Higgs strings~\cite{Vincent:1997cx,Moore:2001px,Bevis:2006mj,Bevis:2007gh,Mukherjee:2010ve,Bevis:2010gj,Hiramatsu:2013tga}, both confirming the approach to the scaling regime.

Recently, cosmic strings have attracted a renewed interest in the context of string
cosmology since it was pointed out that a new type of cosmic strings, \textit{cosmic superstrings}, may be formed at the end of stringy inflation~\cite{Sarangi:2002yt,Jones:2003da,Copeland:2003bj,Dvali:2003zj}.
To our knowledge, the qualitative properties of cosmic superstrings in the late-time universe should be similar to those of field-theoretic strings, except for the fact that the intercommuting probability $P$ is relatively low for cosmic superstrings.
It is normally unity for field-theoretic strings, while it can be significantly smaller than unity for cosmic superstrings.
The authors have extended an analytic description of a network, so-called velocity-dependent one-scale (VOS) model, to include the effect of $P$ in ref.~\cite{Takahashi:2008ui}.
Observables associated with the global properties of a string network, e.g.\ the string number density, are revealed to depend sensitively on the intercommuting probability $P$, and searches for such signals should offer a clue to distinguish cosmic superstrings from field-theoretic strings.

The Gott-Kaiser-Stebbins (GKS) effect \cite{Kaiser:1984iv,Gott:1984ef} is the most characteristic post-recombination effect of a cosmic string in the cosmic microwave background (CMB) sky.
The GKS effect is considered as an integrated Sachs-Wolfe (ISW) effect due to a moving cosmic string, which leads to discontinuities of the CMB temperature fluctuations across the strings with a relative amplitude typically estimated by the dimensionless string tension $G\mu$\,.
The imprint of cosmic strings on the angular power spectrum of the CMB temperature anisotropies have been studied in, e.g., \cite{Hindmarsh:1993pu,Yamauchi:2010ms}, and the current upper bound on the string tension for the strings with $P=1$ is in the range from $1.3\times 10^{-7}$ to $3.2\times 10^{-7}$~\cite{Ade:2013xla}.
Furthermore, cosmic strings generally create non-Gaussian signals
in the CMB temperature anisotropies because topological defects are themselves highly nonlinear objects.
Searches for the string-induced non-Gaussian signals in the CMB may enhance the detectability of cosmic strings, and could not only be used as a tool to prove cosmic strings, but also be helpful to check foreground or systematic contributions.
Non-Gaussian signals induced by the post-recombination effect of a cosmic string network have been estimated in the literature:
references~\cite{Fraisse:2007nu,Takahashi:2008ui,Yamauchi:2010vy} discussed one-point probability distributions of the CMB temperature fluctuations;
also, the CMB temperature bispectrum and trispectrum induced by the GKS effect have been estimated analytically~\cite{Hindmarsh:2009qk,Hindmarsh:2009es,Regan:2009hv,Ringeval:2010ca} and numerically~\cite{Ade:2013xla}.

In this paper, we will study the effect of the weak gravitational lensing by cosmic strings on the CMB temperature anisotropies.
Gravitational lensing by a cosmic string have also been previously studied in the literature~\cite{Yamauchi:2011cu,Yamauchi:2012bc,Yoo:2012dn,Bernardeau:2000xu,Uzan:2000xv,Namikawa:2013wda}.
An observationally important feature is that the lensing events lead to deviations from Gaussianity because a lensed fluctuation is a nonlinear function of fields.
It is known that, in the presence of the cross correlation between the post-recombination CMB temperature fluctuations and the lensing potential, non-vanishing bispectra, which we call the ISW-lensing bispectra hereafter, will appear even if the unlensed temperature fluctuations are exactly Gaussian~\cite{Hu:2000ee}.
As a result, we expect the appearance of the ISW-lensing bispectrum induced by a cosmic string network as yet another string-induced CMB temperature bispectrum.

This paper is organized as follows.
In section~\ref{sec:ISW-lensing bispectrum}, we begin by briefly reviewing the derivation of the ISW-lensing bispectrum and apply it to the case where various gravitational sources exist.
In section~\ref{sec:String-induced bispectra and their detectability}, we introduce the ISW effect due to a cosmic string, namely GKS effect, and the lensing potential due to a cosmic string.
Then we explicitly calculate the string-induced bispectra based on a simple analytic model.
Based on the formulae, prospects for measuring the string-induced CMB temperature bispectra are discussed.
Finally section~\ref{sec:Summary} is devoted to summary and conclusion.

Throughout this paper, we focus on the small patch of sky and work in the flat-sky approximation.
We use the two-dimensional Fourier transformation defined as
%-----------------------------------------------------------------------%
\al{
	f({\bm\theta})
		=\int\frac{\dd^2{\bm\ell}}{(2\pi )^2}f({\bm\ell})e^{i{\bm\ell}\cdot{\bm\theta}}
	\,.\label{eq:Fourier trsf}
}
%-----------------------------------------------------------------------%
where we use the bold letters to label two-vectors on the sky.
Here ${\bm\theta}$ and ${\bm\ell}$ denote the two-dimensional observed position on the sky
and the two dimensional Fourier modes, respectively.
Inner products of two-dimensional vectors are denoted as $ \boldsymbol\theta_1 \cdot \boldsymbol\theta_2 $\,.
We assume a flat $\Lambda$CDM cosmological model as a background
spacetime with the cosmological parameters : $\Omega_{\rm b} h^2=0.22\,,\Omega_\rmm h^2=0.13\,,
\Omega_\Lambda =0.72\,, h=0.7\,, n_\rms =0.96\,, \Delta_\Phi^2 (k=0.002{\rm Mpc}^{-1})=2.4\times 10^{-9}$\,.
We will work in the comoving coordinates
\begin{equation}
g_{\mu\nu}\,\mathrm dx^\mu\,\mathrm dx^\nu
= a(\eta)^2\,
  [-\mathrm d\eta^2 + \delta_{ij}\,\mathrm dr^i\,\mathrm dr^j]
= a(\eta)^2\,
  [-\mathrm d\eta^2 + \mathrm d\chi^2 + \chi^2\,\mathrm d\Omega^2]\,,
\end{equation}
where $ (r^1,r^2,r^3) \equiv \vec r $ is the Cartesian coordinates centered on the observer, $ \chi = |\vec r| $ the comoving distance, and $ \mathrm d\Omega^2 $ the line element on the unit $2$-sphere, which is approximated by $ \mathrm d\boldsymbol\theta \cdot \mathrm d\boldsymbol\theta $ on small scales.
The dot is also used to denote inner products of comoving $3$-vectors: $ \vec r_1 \cdot \vec r_2 \equiv \delta_{ij}\,r_1^i\,r_2^j $\,.

%%%%%%%%%%%%%%%%%%%%%%%%%%%%%%%%%%%%%%%%%%%%%%%%%%%%%%%%%%%%%%%%%%%%%%%%%
%=======================================================================%
\section{ISW-lensing bispectrum}
\label{sec:ISW-lensing bispectrum}
%=======================================================================%
%%%%%%%%%%%%%%%%%%%%%%%%%%%%%%%%%%%%%%%%%%%%%%%%%%%%%%%%%%%%%%%%%%%%%%%%%

In this section we discuss the lensing effect on the CMB temperature anisotropies.
The lensed temperature fluctuation in a direction ${\bm\theta}$\,,
$\tilde\Theta ({\bm\theta})$\,, is Fourier transformed according to
%-----------------------------------------------------------------------%
\al{
	\tilde\Theta ({\bm\ell})
		=\int\dd^2{\bm\theta}\,
			\tilde\Theta ({\bm\theta})\,e^{-i{\bm\ell}\cdot{\bm\theta}}
	\,,\label{eq:Fourier coefficients}
}
%-----------------------------------------------------------------------%
where $\tilde\Theta ({\bm\ell})$ represents the Fourier coefficients.
The auto-bispectrum for the lensed temperature anisotropies in the flat sky is defined as
%-----------------------------------------------------------------------%
\al{
	\ave{\tilde\Theta({\bm\ell}_1) \tilde\Theta({\bm\ell}_2) \tilde\Theta({\bm\ell}_3)}
		=(2\pi)^2\,\delta^2_{\rm D}({\bm\ell}_1+{\bm\ell}_2+{\bm\ell}_3)\,B(\ell_1 ,\ell_2 ,\ell_3)
	\,,\label{eq:bispectrum def}
}
%-----------------------------------------------------------------------%
where the angle brackets $\ave{\cdots}$ denote the ensemble average and $\delta_\mathrm D$ is the Dirac delta function.
We then consider that the lensed temperature fluctuations are related to the unlensed temperature fluctuations through $\tilde\Theta ({\bm\theta})=\Theta ({\bm\theta}+{\bm d})$\,, where $\Theta$ and ${\bm d}$ are the unlensed temperature anisotropies and the deflection angle, respectively.
Assuming that the deflection angle is a perturbed quantity, $|{\bm d}|\ll 1$\,, we can expand the lensed temperature fluctuation as
%-----------------------------------------------------------------------%
\al{
	\tilde\Theta({\bm\theta})
		=\Theta({\bm\theta})
			+{\bm d}({\bm\theta})\cdot{\bm\nabla}\Theta({\bm\theta})
			+\mcO({\bm d}^2)
	\,,\label{eq:tilde Theta expand}
}
%-----------------------------------------------------------------------%
where ${\bm\nabla}$ denotes the two-dimensional covariant derivative on the sky.
Hereafter we neglect the higher-order contributions of $\mcO({\bm d}^2)$ in eq.~\eqref{eq:tilde Theta expand}.
This equation implies that the weak gravitational lensing of the CMB can produce the non-Gaussian temperature fluctuations.
The deflection angle is generally characterized by the sum of two terms:
the gradient of the scalar lensing potential $\phi$ (gradient-mode), and the rotation of the pseudo-scalar lensing potential $\varpi$ (curl-mode)~\cite{Stebbins:1996wx,Hirata:2003ka,Namikawa:2011cs,Yamauchi:2012bc,Yamauchi:2013fra,Namikawa:2013wda}:
%-----------------------------------------------------------------------%
\al{
	{\bm d}({\bm\theta})={\bm\nabla}\phi ({\bm\theta})+\left( *{\bm\nabla}\right)\varpi ({\bm\theta})
	\,,\label{eq:deflection angle decomposition}
}
%-----------------------------------------------------------------------%
where $*$ is the $90$-degree rotation operator.
The Fourier coefficients of the lensed temperature anisotropies are obtained by
performing the two-dimensional Fourier transformation according to eq.~\eqref{eq:Fourier coefficients}.
With the help of eq.~\eqref{eq:deflection angle decomposition}, we find
%-----------------------------------------------------------------------%
\al{
	\tilde\Theta ({\bm\ell})
		=\Theta ({\bm\ell})
			-\int\frac{\dd^2{\bm\ell}_1}{(2\pi )^2}\,
				L({\bm\ell},{\bm\ell}_1)\,\Theta({\bm\ell}_1)
}
%-----------------------------------------------------------------------%
with
\al{
	L(\bm\ell,\bm\ell_1)
		=\bigl[\bm\ell_1\cdot\left(\bm\ell-\bm\ell_1\right)\bigr]\,\phi(\bm\ell-\bm\ell_1)
			+\bigl[\left(*\bm\ell_1\right)\cdot\left(\bm\ell-\bm\ell_1\right)\bigr]\,\varpi(\bm\ell-\bm\ell_1)
	\,,
}
%-----------------------------------------------------------------------%
where $\phi ({\bm\ell})$ and $\varpi ({\bm\ell})$ denote the Fourier components of the gradient- and curl-modes of the deflection angle, respectively.
Even if the unlensed temperature fluctuations $\Theta$ is exactly Gaussian, a non-vanishing bispectrum for the lensed temperature fluctuations will appear and its value can be evaluated as~\cite{Hu:2000ee}
%-----------------------------------------------------------------------%
\al{
	B^{\rm lens}(\ell_1,\ell_2,\ell_3 )
		=\ell_{12}\,C_{\ell_1}^{\Theta\phi}\,C_{\ell_2}^{\Theta\Theta}
			+\left({\rm perms}\right)
	\,,
}
%-----------------------------------------------------------------------%
where $\ell_{mn}=-{\bm\ell}_m\cdot{\bm\ell}_n$\,, (perms) denotes the remaining five permutations of $\{{\bm\ell}_1,{\bm\ell}_2,{\bm\ell}_3\}$, and we have defined the auto- and cross-angular power spectra of the unlensed temperature fluctuations and lensing potential as
%-----------------------------------------------------------------------%
\al{
	\ave{X(\bm\ell)Y(\bm\ell')}
		=(2\pi)^2\,\delta^2_{\rm D}(\bm\ell+\bm\ell')\,C_\ell^{XY}
	\,,\label{eq:power spectrum def}
}
%-----------------------------------------------------------------------%
where $X$ and $Y$ take on $\Theta$ and $\phi$\,.
Note that due to the parity symmetry the cross correlation between the temperature fluctuations and the curl-mode of the deflection angle does not appear.

The cross correlation may be calculated for any secondary effect once its relation to the gravitational potential is given.
For an illustrative example, we shall give the explicit expression for the cross correlation between the lensing potential and the ISW effect due to primordial scalar perturbations.
In a standard $\Lambda$CDM universe, the primordial scalar perturbations give a major contribution to
the gradient-mode of the deflection angle.
In the Born approximation, where the lensing effect is evaluated along the unperturbed light path,
the lensing potential due to primordial scalar perturbations, $\phi_{\rm P}$\,, can be
conveniently evaluated in terms of the Bardeen gravitational potential $\Phi$ as
%-----------------------------------------------------------------------%
\al{
	\phi_{\rm P} ({\bm\theta})
		=-2\int^{\cCMB}_0\!\mathrm d\chi\,\frac{\cCMB -\chi}{\cCMB\chi}\Phi (\eta_0 -\chi ,\chi ,{\bm\theta})
	\,,
}
%-----------------------------------------------------------------------%
where $\cCMB$ is the conformal distance at the last scattering surface and $\eta_0$ the conformal time at present.
On the other hand, the ISW effect due to the primordial scalar perturbations
contributes to the temperature anisotropies as
%-----------------------------------------------------------------------%
\al{
	\Theta_{\rm P}({\bm\theta})
		=-2\int^{\cCMB}_0\dd\chi\,\dot\Phi (\eta_0 -\chi ,\chi ,{\bm\theta})
	\,,
}
%-----------------------------------------------------------------------%
where the dot ( $\dot{}$ ) denotes the derivative with respect to the conformal time.
It follows that the flat-sky cross correlation is given by \cite{Hu:2000ee}
%-----------------------------------------------------------------------%
\al{
	C_\ell^{\Theta_{\rm P}\phi_{\rm P}}
		=\frac{2\pi^2}{\ell^3}
			\int^{\cCMB}_0\dd\chi
			\chi\left( -2\dot F(\chi )\right)
			\left( -2F(\chi )\frac{\cCMB -\chi}{\cCMB\chi}\right)
			\Delta^2_\Phi\left(\ell /\chi\right)
	\,,\label{eq:C_l^PP}
}
%-----------------------------------------------------------------------%
where $\Delta_\Phi^2(k)$ is the dimensionless primordial power spectrum of the Bardeen potential and $F(\chi)$ is given by
%-----------------------------------------------------------------------%
\al{
	F(\chi )
		=(1+z)\frac{H(z)}{H_0}
		\frac{\int^\infty_z\dd z' (1+z')H^{-3}(z')}{\int^\infty_0\dd z'' (1+z'')H^{-3}(z'')}
	\,,
}
%-----------------------------------------------------------------------%
where $z$ is the redshift, which is related to the conformal distance through
$\chi=\int_0^z\!\dd z'/H(z')$\,.
We should note that the unlensed temperature power spectrum $C_\ell^{\Theta_{\rm P}\Theta_{\rm P}}$ has a negligibly small amplitude on small scales due to the Silk damping, while the cross-correlation between the temperature and the lensing potential $C_\ell^{\Theta_{\rm P}\phi_{\rm P}}$ could have a non-vanishing amplitude even at small scales.

We then apply the derivation of the ISW-lensing bispectrum, originally developed in the theoretical studies of the bispectrum induced by primordial density perturbations, to the case of various gravitational sources.
To evaluate the various types of bispectra, we first assume that the observed sky map of the temperature anisotropies can be regarded as a superposition of those due to each source for simplicity.
Let us introduce the index $\alpha$ to denote the contribution from each kind of sources as
%-----------------------------------------------------------------------%
\al{
	\tilde\Theta ({\bm\theta})
		=\sum_\alpha\tilde\Theta_\alpha ({\bm\theta})
		=\sum_\alpha\Theta_\alpha\left({\bm\theta}+{\bm\nabla}\phi\right)
	\,.\label{eq:Theta_alpha def}
}
%-----------------------------------------------------------------------%
Here we have assumed that the deflection angle can be described only by the gradient-mode of the the deflection angle although the curl-mode in general contributes the deflection angle (see eq.~\eqref{eq:deflection angle decomposition} and, e.g., refs.~\cite{Namikawa:2011cs,Yamauchi:2012bc,Yamauchi:2013fra,Namikawa:2013wda} for the estimation of the string-induced curl mode).
Similarly, since the gradient mode of the deflection angle strongly depends on the source gravitational potential and its distribution, we assume that the total scalar lensing potential can be decomposed into each kind of contributions as
%-----------------------------------------------------------------------%
\al{
	\phi ({\bm\theta})
		=\sum_{\alpha}\phi_\alpha ({\bm\theta})
	\,.\label{eq:phi_beta def}
}
%-----------------------------------------------------------------------%
Using eqs.~\eqref{eq:Theta_alpha def} and \eqref{eq:phi_beta def} and expanding the lensed temperature anisotropies up to $\mcO(\phi)$, we can rewrite it as
%-----------------------------------------------------------------------%
\al{
	\tilde\Theta ({\bm\theta})
		=\sum_\alpha\Theta_\alpha\left({\bm\theta}+\sum_\beta{\bm\nabla}\phi_\beta\right)
		=\sum_\alpha
			\left(
				\Theta_\alpha ({\bm\theta})
				+\sum_{\beta}{\bm\nabla}\phi_\beta ({\bm\theta})\cdot{\bm\nabla}\Theta_\alpha ({\bm\theta})
			\right)
	\,.
}
%-----------------------------------------------------------------------%
Hence the flat-sky bispectrum for the lensed temperature anisotropies, eq.~\eqref{eq:bispectrum def}, can be decomposed as
%-----------------------------------------------------------------------%
\al{
	B (\ell_1 ,\ell_2 ,\ell_3 )
		=\sum_\alpha B^{\alpha\alpha\alpha}( \ell_1 ,\ell_2 ,\ell_3 )
			+\sum_{\alpha ,\beta}B^{\alpha\beta}(\ell_1 ,\ell_2 ,\ell_3 )
}
%-----------------------------------------------------------------------%
where we have introduced $B^{\alpha\alpha\alpha}$ to denote the bispectrum for the unlensed temperature anisotropies generated by the gravitational source $\alpha$\,, and $B^{\alpha\beta}$ to denote {\it the $\alpha\beta$-type ISW-lensing bispectrum}, which is defined by
%-----------------------------------------------------------------------%
\al{
	B^{\alpha\beta}(\ell_1 ,\ell_2 ,\ell_3 )
		=\ell_{12}\,C_{\ell_1}^{\Theta_\alpha\phi_\alpha}C_{\ell_2}^{\Theta_\beta\Theta_\beta}
			+\left({\rm perms}\right)
	\,,
}
%-----------------------------------------------------------------------%
with $\ell_{mn}=-{\bm\ell}_m\cdot{\bm\ell}_n$\,.
Here we have neglected the connected part of the four-point function of
the temperature fluctuations and the lensing potential for simplicity.

In this paper we focus only on the contributions from inflationary primordial fluctuations (P) and a cosmic string network (S).
The
% leading part of the
resultant bispectrum for the lensed temperature anisotropies can be decomposed as
%-----------------------------------------------------------------------%
\al{
	B=B^{\rm PPP}+B^{\rm PP}+B^{\rm SSS}+B^{\rm SS}+B^{\rm SP}+B^{\rm PS}
	\,.\label{eq:B decomposition}
}
%-----------------------------------------------------------------------%
The first two terms in eq.~\eqref{eq:B decomposition}\,, $B^{\rm PPP}$ and $B^{\rm PP}$\,, correspond to the standard unlensed and ISW-lensing bispectra (see eq.~\eqref{eq:C_l^PP}) due to the primordial scalar perturbations, respectively.
A recent observation \cite{Ade:2013ydc,Ade:2013dsi} shows that there is yet no evidence for any primordial non-Gaussianity, but the ISW-lensing bispectrum expected in the standard $\Lambda$CDM universe has been measured at more than $2\sigma$ statistical significance.
On the other hand, taking into account the contributions from a string network, we have four additional components;
$B^{\rm SSS}$ represents the bispectrum purely due to the GKS effect of strings (see next section for the GKS effect), which has been estimated in the literature~\cite{Hindmarsh:2009qk,Hindmarsh:2009es,Regan:2009hv,Ringeval:2010ca,Ade:2013xla}, whereas the new types of the string-induced, ISW-lensing, bispectra $B^{\rm SS}\,, B^{\rm SP}$ and $B^{\rm PS}$ have appeared through the CMB lensing.
The remainder of the paper will be devoted to the evaluation of these new bispectra.

Before closing this section, we should discuss possible modifications to eq.~\eqref{eq:B decomposition} from the string-induced non-Gaussian correlations.
Since each photon scattering by a cosmic string produces strongly non-Gaussian signals in the CMB, the connected part of the four-point functions such as $\ave{\Theta_{\rm S}\Theta_{\rm S}\Theta_{\rm S}\phi_{\rm S}}$ and higher-order correlation functions would give non-vanishing contributions in eq.~\eqref{eq:B decomposition}.
However, $\Theta_{\rm S}$ can be actually treated as nearly Gaussian variable and these modifications should be small.
This is because a photon ray is scattered by cosmic strings many times through its way from the last scattering surface to an observer.
Hence $\Theta_{\rm S}$ would behave like a random walk and its probability distribution function may be approximated by a Gaussian distribution~\cite{Takahashi:2008ui,Yamauchi:2010vy}.
Although we ignore those small non-Gaussian modifications hereafter, they would rather enhance the signals, and the expected signal-to-noise ratios will be increased.
In this sense, the analysis we will give later would give a rather conservative estimate for the detectability of cosmic strings.

%%%%%%%%%%%%%%%%%%%%%%%%%%%%%%%%%%%%%%%%%%%%%%%%%%%%%%%%%%%%%%%%%%%%%%%%%
%=======================================================================%
\section{String-induced bispectra and their detectability}
\label{sec:String-induced bispectra and their detectability}
%=======================================================================%
%%%%%%%%%%%%%%%%%%%%%%%%%%%%%%%%%%%%%%%%%%%%%%%%%%%%%%%%%%%%%%%%%%%%%%%%%

In this section, we consider the ISW effect and the gravitational lensing due to a cosmic string network as yet another source of the CMB temperature bispectrum.
After briefly reviewing the properties of the post-recombination effect of the cosmic string, namely the Gott-Kaiser-Stebbins (GKS) effect and the string-induced lensing potential, in section~\ref{sec:GKS effect and string lensing}, we give the explicit expression for the string-induced bispectra in section~\ref{sec:String correlations}.
The signal-to-noise ratios for the string-induced bispectra are estimated,
and the detectability of the string network is discussed in section \ref{sec:Signal-to-noise ratio}.

%%%%%%%%%%%%%%%%%%%%%%%%%%%%%%%%%%%%%%%%%%%%%%%%%%%%%%%%%%%%%%%%%%%%%%%%%
%=======================================================================%
\subsection{GKS effect and string lensing}
\label{sec:GKS effect and string lensing}
%=======================================================================%
%%%%%%%%%%%%%%%%%%%%%%%%%%%%%%%%%%%%%%%%%%%%%%%%%%%%%%%%%%%%%%%%%%%%%%%%%

We first consider the GKS effect as an ISW effect due to a cosmic string.
The ISW formula is given by
%-----------------------------------------------------------------------%
\al{
	\Theta_{\rm S} ({\bm\theta})
		=-\frac{1}{2}\int^{\cCMB}_0\dd\chi\frac{\dd x^\mu}{\dd\chi}\frac{\dd x^\nu}{\dd\chi}
			\dot h_{\mu\nu}(\eta_0 -\chi ,\chi ,{\bm\theta})
	\,,
}
%-----------------------------------------------------------------------%
where $h_{\mu\nu}$ is the metric perturbation caused by strings and $\dd x^\mu /\dd\chi =(-1,\vec n)$ denotes the null vector along the line of sight with $\vec n$ being a unit vector pointing the photon propagation direction in the background spacetime.
In order to evaluate the metric perturbations through the linearized Einstein equations, we write down the string stress-energy tensor.
To do so, we assume that a string segment can be well approximated as a Nambu-Goto string and we introduce the three-dimensional embedding function of string position as $\vec r=\vec r(\sigma,\eta)$\,, where $\sigma$ is the spacelike worldsheet coordinate.
The stress-energy tensor for a string segment in the transverse gauge is described as
%-----------------------------------------------------------------------%
\al{
	T^{\mu\nu}(\eta,{\vec r})
		=\mu\int\dd\sigma
			\left(
			\begin{array}{cc}
				1 & -\dot r^i\\
				-\dot r^j & \dot r^i\,\dot r^j -r^i{}'\,r^j{}'\\
			\end{array}
			\right)
			\delta_\rmD^3 (\vec r-\vec r(\sigma ,\eta ))
	\,,\label{eq:string stress-energy}
}
%-----------------------------------------------------------------------%
where $\mu$ is the string tension, the dot ( $\dot{}$ ) and the prime ( ${}'$ ) denote the derivatives with respect to $\eta$ and $\sigma$\,, respectively.

The stress-energy tensor should be properly evaluated along the line of sight, on which $\eta=\eta_0-\chi$\,.
To do so in an analytical manner, we further impose that the string segment as seen by the observer is localized at a certain redshift, namely the distance on the lightcone between the observer and the string segment can be well approximated by a constant value, $|{\vec r}(\sigma,\eta_0-\chi)|\approx\chi_{\rm S} ={\rm const}$ \cite{Yamauchi:2011cu}.
This condition is solved for the conformal distance as $\chi=\chi(\sigma;\chi_\mathrm S)$\,, so that we can parameterize the string position as seen by the observer as $\vec r=\vec r_\mathrm S(\sigma)\equiv\vec r(\sigma,\eta_0-\chi(\sigma;\chi_\mathrm S))$\,.
With this approximation, let us define the two-dimensional angular position of a string $ \boldsymbol\theta_\mathrm S $ by
%-----------------------------------------------------------------------%
\al{
	{\bm\theta}_{\rm S}(\sigma)
		\equiv\frac{1}{\chi_{\rm S}}
				\left(
					\vec e_1 \cdot \vec r_\mathrm S(\sigma)\,,\ 
					\vec e_2 \cdot \vec r_\mathrm S(\sigma)
				\right)
	\,,
}
%-----------------------------------------------------------------------%
where the orthogonal projectors $ (\vec e_1, \vec e_2) $ satisfy $ \vec e_a \cdot \vec e_b = \delta_{ab}\,, \vec e_a \cdot \vec n = 0 $\,.
Similarly, the angular velocity $ \dot{\boldsymbol\theta}_\mathrm S(\sigma) $ is defined by replacing $ r_\mathrm S^i(\sigma) $ with $ \dot r_\mathrm S^i(\sigma) \equiv \dot r^i(\sigma,\eta_0-\chi(\sigma;\chi_\mathrm S)) $ in the right-hand side of the above expression for $ \boldsymbol\theta_\mathrm S(\sigma) $\,.

%>>>>>>>>>>>>>>>>>>>>>>>>>>>>>>>FIGURE<<<<<<<<<<<<<<<<<<<<<<<<<<<<<<<<<<%
\begin{figure}[tbp]
\bc
\includegraphics[width=80mm]{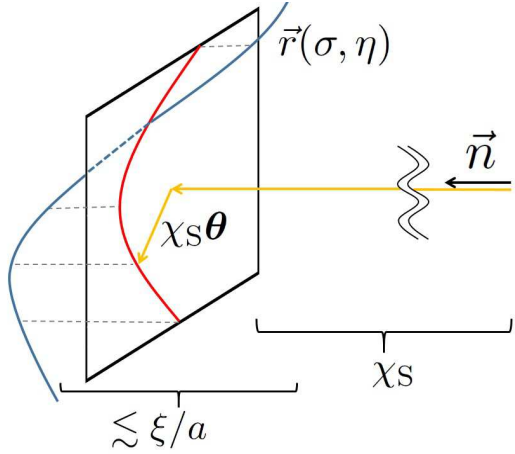}
\caption{
A representation of the two-dimensional observed position $\bm\theta$\,, and the comoving distance between the string segment and the observer $\chi_{\rm S}$\,, the three-dimensional embedding function of the string position $\vec r(\sigma,\eta)$\,, and the projected string angular position $\bm\theta_{\rm S}(\sigma)$\,.
}
\label{setup}
\ec
\end{figure}
%>>>>>>>>>>>>>>>>>>>>>>>>>>>>>>>>>><<<<<<<<<<<<<<<<<<<<<<<<<<<<<<<<<<<<<%
In figure~\ref{setup}\,, we show the representation of the quantities used in this paper.
The above approximation should be valid as long as we focus on distant strings at small patch of sky;
since the correlation length of a string segment $\xi$ (see figure~\ref{setup}) is known to grow in proportion to the Hubble length, the extension of a string along the line of sight is bounded by $1/H$\,, so it should be much smaller than the physical distance.

Under this approximation, the stress-energy tensor of a string along the line of sight is described as~\cite{Yamauchi:2011cu,Bernardeau:2000xu,Uzan:2000xv}
%-----------------------------------------------------------------------%
\al{
	\frac{\dd x^\mu}{\dd\chi}\frac{\dd x^\nu}{\dd\chi}
		T_{\mu\nu}(\eta_0 -\chi ,\chi ,{\bm\theta})
		\approx\frac{\mu}{\chi_{\rm S}^2}\,\delta_{\rm D}(\chi -\chi_{\rm S})
			\int\dd\sigma\,
			\delta_{\rm D}^2(\boldsymbol\theta-\boldsymbol\theta_\mathrm S(\sigma))
	\,.
}
%-----------------------------------------------------------------------%
With these notations, the temperature fluctuations due to the GKS effect
is evaluated by the following formula~\cite{Hindmarsh:2009qk,Hindmarsh:2009es,Regan:2009hv,Ringeval:2010ca,Hindmarsh:1993pu}:
%-----------------------------------------------------------------------%
\al{
	{\bm\nabla}^2\Theta_{\rm S} ({\bm\theta})
		&=8\pi G\,
			\int^{\cCMB}_0\dd\chi\,\chi^2
			\frac{\dd x^\mu}{\dd\chi}\frac{\dd x^\nu}{\dd\chi}
			\dot T_{\mu\nu}(\eta_0 -\chi ,\chi ,{\bm\theta})
	\notag\\
		&\approx 8\pi G\mu\,
			\int\dd\sigma\,
			(\dot{\bm\theta}_{\rm S}(\sigma)\cdot{\bm\nabla})
			\delta_{\rm D}^2({\bm\theta} -{\bm\theta}_{\rm S}(\sigma))
	\,,\label{eq:GKS in real space}
}
%-----------------------------------------------------------------------%
where we have used \Blue{the} linearized Einstein equations in the first line of eq.~\eqref{eq:GKS in real space}.
Performing the two-dimensional Fourier transformation, we obtain the Fourier coefficients of the GKS temperature fluctuations as
%-----------------------------------------------------------------------%
\al{
	\Theta_{\rm S} ({\bm\ell})
		=i8\pi G\mu\frac{1}{\ell^2}\,
			\int\dd\sigma\,
			({\bm\ell}\cdot\dot{\bm\theta}_{\rm S}(\sigma))\,
				e^{-i{\bm\ell}\cdot{\bm\theta}_{\rm S}(\sigma)}
	\,.
	\label{eq:GKS effect}
}
%-----------------------------------------------------------------------%

On the other hand, the lensing potential induced by a cosmic string, $\phi_{\rm S}$\,, 
are related to the convergence field $\kappa_{\rm S}$ through $\kappa_{\rm S}=\boldsymbol{\nabla}^2\phi_{\rm S}/2$\,.
The convergence field can be described in terms of the perturbed Ricci tensor
and is related to the stress-energy tensor through the perturbed
Einstein equations as \cite{Yamauchi:2011cu}
%-----------------------------------------------------------------------%
\al{
	\kappa_{\rm S} ({\bm\theta})
		=4\pi G\int^{\cCMB}_0\dd\chi\frac{(\cCMB -\chi )\chi}{\cCMB}
			\frac{\dd x^\mu}{\dd\chi}\frac{\dd x^\nu}{\dd\chi}
			T_{\mu\nu}(\eta_0 -\chi ,\chi ,{\bm\theta})
	\,.
}
%-----------------------------------------------------------------------%
Assuming that the string segment is localized at a certain redshift, as we mentioned above, the above expression can be reduced to
%-----------------------------------------------------------------------%
\al{
	\boldsymbol{\nabla}^2\phi_{\rm S}({\bm\theta})
		\approx 8\pi G\mu\frac{\cCMB -\chi_{\rm S}}{\cCMB\,\chi_{\rm S}}\,
				\int\dd\sigma\,
					\delta_{\rm D}^2({\bm\theta}-{\bm\theta}_{\rm S}(\sigma))
	\,.
}
%-----------------------------------------------------------------------%
Hence we obtain the Fourier coefficients of the string lensing potential as
%-----------------------------------------------------------------------%
\al{
	\phi_{\rm S} ({\bm\ell})
		=-8\pi G\mu\frac{\cCMB -\chi_{\rm S}}{\cCMB\,\chi_{\rm S}}\frac{1}{\ell^2}\,
			\int\dd\sigma\,e^{-i{\bm\ell}\cdot{\bm\theta}_{\rm S}(\sigma)}
	\,.
	\label{eq:scalar lensing potential}
}
%-----------------------------------------------------------------------%
For a cosmic string network, non-vanishing cross-correlation 
is expected to exist between the GKS temperature fluctuations 
and the string-induced lensing potential.

%%%%%%%%%%%%%%%%%%%%%%%%%%%%%%%%%%%%%%%%%%%%%%%%%%%%%%%%%%%%%%%%%%%%%%%%%
%=======================================================================%
\subsection{String correlations}
\label{sec:String correlations}
%=======================================================================%
%%%%%%%%%%%%%%%%%%%%%%%%%%%%%%%%%%%%%%%%%%%%%%%%%%%%%%%%%%%%%%%%%%%%%%%%%

Before going into the details, let us discuss the dependence of the string-induced bispectra on the string tension $G \mu$\,.
From the expressions of $\Theta_\mathrm S(\boldsymbol\ell)$ and $\phi_\mathrm S(\boldsymbol\ell)$\,, eqs.~\eqref{eq:GKS effect} and \eqref{eq:scalar lensing potential}, we deduce that the power spectra $C_\ell^{XY}\propto\langle X\,Y\rangle$ (for $X,Y=\Theta_\mathrm S,\phi_\mathrm S$) scale as $\propto(G \mu)^2$\,.
Therefore we find the following proportionalities of the string-induced ISW-lensing:
\begin{equation}
B^{\mathrm S\mathrm S}
\propto
  (G \mu)^4\,,
\quad
B^{\mathrm S\mathrm P}
\propto
  (G \mu)^2\,,
\quad
B^{\mathrm P\mathrm S}
\propto
  (G \mu)^2\,.
\end{equation}
On the other hand, the purely GKS-induced bispectrum obeys $B^{\rm SSS}\propto (G\mu)^3$ \cite{Hindmarsh:2009qk,Regan:2009hv,Ringeval:2010ca,Ade:2013xla}.
Hence, in the case of the smaller string tension, the SP- and PS-type contributions could dominate the total bispectrum rather than the SSS-type.
Moreover, at small scales where the unlensed primordial fluctuations are damped, the unlensed primordial spectrum $C_\ell^{\Theta_{\rm P}\Theta_{\rm P}}$ has little power and only the cross-correlation $C_\ell^{\Theta_{\rm P}\phi_{\rm P}}$ is relevant.
Therefore, we expect that the SP-type bispectrum, which obeys the proportionality $B^{\rm SP}\propto C_{\ell_1}^{\Theta_{\rm S}\phi_{\rm S}}C_{\ell_2}^{\Theta_{\rm P}\Theta_{\rm P}}$\,, is exponentially small at small scales whereas the PS-type, $B^{\rm PS}\propto C_{\ell_1}^{\Theta_{\rm P}\phi_{\rm P}}C_{\ell_2}^{\Theta_{\rm S}\Theta_{\rm S}}$\,, should give the most significant contributions.
According to the above observations, we shall only consider $B^{\mathrm S \mathrm P}$ and $B^{\mathrm P \mathrm S}$ in what follows.

For our purpose, we need to evaluate $ C_\ell^{\Theta_\mathrm S\Theta_\mathrm S} $ and $ C_\ell^{\Theta_\mathrm S\phi_\mathrm S} $\,.
In order to calculate the angular power spectrum, we follow the analytic approach~\cite{Yamauchi:2010ms}, originally developed in the studies of the Sunyaev-Zel'dovich effect~\cite{Komatsu:2002wc,Komatsu:1999ev,Cole:1989vx}.
The GKS fluctuations and the string-induced lensing potential are characterized by the distance $\chi_{\rm S}$ and the parameters for the string-segment configuration $\{\psi_a\}\,(a=1,2,\cdots)$ including the set of the angular parameters for the string directions and the curvature.
The observed sky maps of the temperature fluctuations and the lensing potential are assumed to appear as a superposition of each contribution, namely $\Theta_{\rm S}^{\rm tot}({\bm\ell})=\sum_i\Theta_{\rm S}({\bm\ell};\chi_{{\rm S},i},\{\psi_{i,a}\})$\,, $\phi_{\rm S}^{\rm tot}({\bm\ell})=\sum_i\phi_{\rm S}({\bm\ell};\chi_{{\rm S},i},\{\psi_{i,a}\})$\,, with ``$i$'' denoting the contribution from the $i$-th string segment.
The angular power spectrum then can be decomposed into two parts:
the contributions from the Poisson-distributed string segments and those from the correlations between the different segments.
At small scales, the angular power spectrum will be dominated by the contribution of the sum of statistically independent segments even if the segment-segment correlation is taken into account.
With the help of eqs.~\eqref{eq:power spectrum def}\,, \eqref{eq:GKS effect}\,, and \eqref{eq:scalar lensing potential}\,, we find
%-----------------------------------------------------------------------%
\al{
	C_\ell^{\Theta_{\rm S}\Theta_{\rm S}}
		&=\frac{1}{\mcA}\ave{\Theta_{\rm S}^{\rm tot}({\bm\ell})\Theta_{\rm S}^{\rm tot}(-{\bm\ell})}
	\notag\\
		&\approx \frac{(8\pi G\mu )^2}{\mcA}
			\frac{1}{\ell^4}
			\int^{\chi_{\rm CMB}}_0\dd\chi\,\frac{\dd V}{\dd\chi}\,
			\left(\prod_a\int\!\dd\psi_a\right)\,f_\mathrm S(\{\psi_a\})
	\notag\\
	&\quad\quad\times
					\int\dd\sigma_1\dd\sigma_2
					({\bm\ell}\cdot\dot{\bm\theta}_{\rm S}(\sigma_1))\,
					({\bm\ell}\cdot\dot{\bm\theta}_{\rm S}(\sigma_2))\,
								e^{i{\bm\ell}\cdot({\bm\theta}_{\rm S}(\sigma_1)-{\bm\theta}_{\rm S}(\sigma_2))}
	\label{eq:Cl_Theta_STheta_S}
}
%-----------------------------------------------------------------------%
for the angular power spectrum for the GKS temperature anisotropies, and
%-----------------------------------------------------------------------%
\al{
	C_\ell^{\Theta_{\rm S}\phi_{\rm S}}
		&=\frac{1}{\mcA}\ave{\Theta_{\rm S}^{\rm tot}({\bm\ell})\phi_{\rm S}^{\rm tot}(-{\bm\ell})}
	\notag\\
		&\approx -i\frac{(8\pi G\mu )^2}{\mcA}\frac{1}{\ell^4}
			\int^{\chi_{\rm CMB}}_0\dd\chi\,\frac{\dd V}{\dd\chi}
			\frac{\cCMB -\chi}{\cCMB\,\chi}\,
			\left(\prod_a\int\!\dd\psi_a\right)\,f_\mathrm S(\{\psi_a\})
	\notag\\
	&\quad\quad\times
				\int\dd\sigma_1\,\dd\sigma_2\,
					({\bm\ell}\cdot\dot{\bm\theta}_{\rm S}(\sigma_1))\,
					e^{i{\bm\ell}\cdot(\bm\theta_{\rm S}(\sigma_1)-\bm\theta_{\rm S}(\sigma_2))}
	\label{eq:Cl_Theta_Sphi_S}
}
%-----------------------------------------------------------------------%
for the cross correlation between the GKS anisotropies and the string-induced lensing potential, where $\mcA=(2\pi)^2\,\delta^2_{\rm D}({\bm 0})=4\pi\,f_{\rm sky}$ is the area size with $f_{\rm sky}$ being the fractional sky coverage, $(\dd V/\dd\chi)\,\dd\chi=4\pi\chi^2\,\dd\chi$ and $(\prod_a\dd\psi_a)\,f_\mathrm S(\{\psi_a\})=\mathrm dn_\mathrm S(\{\psi_a\})$ denote the comoving differential volume element at a distance $\chi$ and the comoving number density of string segments with the string configuration parameters in the range $[\psi_a,\psi_a+\dd\psi_a]$\,, respectively.
It is in general difficult to evaluate the average for the string configuration parameters, though we can calculate it explicitly when we focus on the exactly straight string-segments~\cite{Yamauchi:2010ms}.

Instead, we will use the simple analytic model to estimate the correlations within the string segment developed by \cite{Hindmarsh:1993pu,Vincent:1996qr,Albrecht:1997mz}.
In this model, the notion of the string-segment configuration average $\ave{\cdots}_{\rm seg}$ is introduced (to be distinguished from the usual meaning of the ensemble average $\ave{\cdots}$), which allows evaluation of the integration over the string configuration parameters through the correspondence $[\prod_a(\int\!\dd\psi_a)\,f_\mathrm S(\{\psi_a\})\,\cdots]\rightarrow n_{\rm S}\,\ave{\cdots}_{\rm seg}$\,, where $n_\mathrm S$ is the comoving number density of the strings.
Furthermore, the variables $r^i{}'$ and $\dot r^i$ are assumed to be exactly Gaussian and isotropic with mean zero, and all the equal-time correlations can be expressed in terms of the following two point functions:
%-----------------------------------------------------------------------%
\al{
	&\ave{\dot r^i(\sigma_1,\eta)\,\dot r^j(\sigma_2,\eta)}_{\rm seg}
		=\frac{1}{3}\delta^{ij}\,V_{\rm S}(\sigma_1-\sigma_2,\eta)
	\,,\label{eq:V_S def}\\
	&\ave{r^i{}'(\sigma_1,\eta)\,r^j{}'(\sigma_2,\eta)}_{\rm seg}
		=\frac{1}{3}\delta^{ij}\,T_{\rm S}(\sigma_1-\sigma_2,\eta)
	\,,\label{eq:T_S def}\\
	&\ave{r^i{}'(\sigma_1,\eta)\,\dot r^j(\sigma_2,\eta)}_{\rm seg}
		=\frac{1}{3}\delta^{ij}\,M_{\rm S}(\sigma_1-\sigma_2,\eta)
	\,.\label{eq:M_S def}
}
%-----------------------------------------------------------------------%
Then the asymptotic forms of these correlators\,, $V_{\rm S}$\,, $T_{\rm S}$\,, and $M_{\rm S}$ are estimated based on the velocity dependent one-scale (VOS) model~\cite{Martins:2000cs,Martins:1996jp,Avgoustidis:2005nv}.
In VOS, a string network is characterized by the correlation length $\xi =1/(H\gamma_{\rm S})$ and the root-mean-square velocity $v_{\rm rms}$\,.
Taking into account the probabilistic nature of the intercommuting process, for $P\ll 1$, we obtain the approximate expressions $\gamma_{\rm S}\approx [\pi\sqrt{2}/(3\tilde cP)]^{1/2}\approx 2.5 (\tilde cP/0.23)^{-1/2}$ and $v_{\rm rms}^2\approx (1/2)\,[1-\pi /(3\gamma_{\rm S})]$ in the matter-dominated era \cite{Takahashi:2008ui}, where $\tilde c\approx 0.23$ quantifies the efficiency of the loop formation~\cite{Martins:2000cs}, and $P$ is the intercommuting probability.
Since in our calculation we consider only a string segment with length $\sim\xi$\,, the correlators in eqs.~\eqref{eq:V_S def}-\eqref{eq:M_S def} are expected to be damped on scales larger than the correlation length of the string network, that is for $|\sigma|\gtrsim\xi /a$\,, while have the non-vanishing expectation values for $|\sigma |\lesssim \xi /a$\,.
In terms of the scaling quantities of the string network, namely $\xi$ and $v_{\rm rms}$\,, the asymptotic behaviors of the two-point correlators are given by (see e.g.\ \cite{Hindmarsh:1993pu})
%-----------------------------------------------------------------------%
\al{
	&V_{\rm S} (\sigma ,\eta )=
		\Biggl\{
			\begin{array}{ll}
			v_{\rm rms}^2\  &\  (\sigma\lesssim\xi/a)\\
			0\  &\  (\sigma\gtrsim\xi/a)\\
			\end{array}
	\,,\label{eq:asymptotic value of V_S}\\
	&T_{\rm S} (\sigma ,\eta )=
		\Biggl\{
			\begin{array}{ll}
			1-v_{\rm rms}^2&\  (\sigma\lesssim\xi/a)\\
			0\  &\  (\sigma\gtrsim\xi/a)\\
			\end{array}
	\,,\\
	&M_{\rm S} (\sigma ,\eta )=
		\Biggl\{
			\begin{array}{ll}
			c_0\,a\,\sigma/\xi\  &\  (\sigma\lesssim\xi/a)\\
			0\  &\  (\sigma\gtrsim\xi/a)\\
			\end{array}
	\,,\label{eq:asymptotic value of Pi_S}
}
%-----------------------------------------------------------------------%
where $c_0=(\xi/a)\,\ave{\dot{\vec r}\cdot\vec r{}''}_{\rm seg}$ represents 
the cross correlator between the string velocity and curvature.
The non-vanishing cross correlation $M_\mathrm S$ appears in the cosmological background, while one can see it vanishes in a flat spacetime~\cite{Yamauchi:2010vy,Hindmarsh:2009qk}.
In the scaling regime, $c_0$ can be evaluated in terms of the root-mean-square velocity through the VOS scaling equations as
$c_0=(2\sqrt{2}/\pi )\,v_{\rm rms}\,(1-v_{\rm rms}^2)\,(1-8v_{\rm rms}^6)/(1+8v_{\rm rms}^6)\approx [3\tilde cP/(\pi\sqrt{2})]^{1/2}/2\approx 0.19(\tilde cP/0.23)^{1/2}$\,.

Consequently, in the model described above, eqs.~\eqref{eq:Cl_Theta_STheta_S} and \eqref{eq:Cl_Theta_Sphi_S} can be rewritten as
%-----------------------------------------------------------------------%
\al{
	C_\ell^{\Theta_{\rm S}\Theta_{\rm S}}
		\approx &\frac{(8\pi G\mu )^2}{\mcA}\,
			\frac{1}{\ell^4}\,
			\int^{\chi_{\rm CMB}}_0\dd\chi\,\frac{\dd V}{\dd\chi}\,n_{\rm S}
	\notag\\
	&\quad\times
					\int\dd\sigma_1\,\dd\sigma_2\,
					\ave{
						(\bm\ell\cdot\dot{\bm\theta}_{\rm S}(\sigma_1))\,
						(\bm\ell\cdot\dot{\bm\theta}_{\rm S}(\sigma_2))\,
						e^{i\bm\ell\cdot(\bm\theta_{\rm S}(\sigma_1)-\bm\theta_{\rm S}(\sigma_2))}
					}_{\rm seg}
	\,,\label{eq:Cl_Theta_STheta_S 2}\\
	C_\ell^{\Theta_{\rm S}\phi_{\rm S}}
		\approx &-i\frac{(8\pi G\mu )^2}{\mcA}\,\frac{1}{\ell^4}\,
			\int^{\chi_{\rm CMB}}_0\dd\chi\,\frac{\dd V}{\dd\chi}\,n_{\rm S}\,
			\frac{\cCMB-\chi}{\cCMB\,\chi}
	\notag\\
	&\quad\times
				\int\dd\sigma_1\,\dd\sigma_2\,
				\ave{
					(\bm\ell\cdot\dot{\bm\theta}_{\rm S}(\sigma_1))\,
					e^{i\bm\ell\cdot(\bm\theta_{\rm S}(\sigma_1)-\bm\theta_{\rm S}(\sigma_2))}
				}_{\rm seg}
	\,.\label{eq:Cl_Theta_Sphi_S 2}
}
%-----------------------------------------------------------------------%
The comoving string number density can be estimated in terms of the correlation length $\xi$ as $n_{\rm S}\approx a^3/\xi^3=a^3H^3\gamma_{\rm S}^3$\,.
By virtue of the properties of the string correlators in eqs.~\eqref{eq:V_S def}-\eqref{eq:M_S def}, and recalling that the distant strings can be treated as thin objects, we can evaluate the string segment configuration averages as
%-----------------------------------------------------------------------%
\al{
	&\ave{
		(\bm\ell\cdot\dot{\bm\theta}_{\rm S}(\sigma_1))\,
		(\bm\ell\cdot\dot{\bm\theta}_{\rm S}(\sigma_2))\,
		e^{i\bm\ell\cdot(\bm\theta_{\rm S}(\sigma_1)-\bm\theta_{\rm S}(\sigma_2))}
	}_{\rm seg}
	\notag\\
	&\quad
			\approx\frac{1}{3}\frac{\ell^2}{\chi_\mathrm S^2}\,
			\biggl\{
				V_{\rm S}(\sigma_1-\sigma_2,\eta_0-\chi_{\rm S})
				-\frac{1}{3}\ell^2\Pi_{\rm S}^2(\sigma_1-\sigma_2 ,\eta_0-\chi_{\rm S})
			\bigg\}
        \notag\\
        &\quad\quad
			\times\exp\biggl[-\frac{1}{6}\ell^2\Gamma_{\rm S}(\sigma_1 -\sigma_2,\eta_0-\chi_{\rm S})\biggr]
	\,,\\
	&\ave{
		(\bm\ell\cdot\dot{\bm\theta}_{\rm S}(\sigma_1))\,
		e^{i\bm\ell\cdot(\bm\theta_{\rm S}(\sigma_1)-\bm\theta_{\rm S}(\sigma_2))}
	}_{\rm seg}
	\notag\\
	&\quad
		\approx\frac{i}{3}\frac{\ell^2}{\chi_\mathrm S}\,\Pi_{\rm S}(\sigma_1-\sigma_2,\eta_0-\chi_{\rm S})\,
			\exp\biggl[-\frac{1}{6}\ell^2\Gamma_{\rm S}(\sigma_1 -\sigma_2,\eta_0-\chi_{\rm S})\biggr]
	\,,
}
%-----------------------------------------------------------------------%
where we have introduced $\Gamma_{\rm S}$ and $\Pi_{\rm S}$ defined by
%-----------------------------------------------------------------------%
\al{
	&\Gamma_{\rm S}(\sigma_1 -\sigma_2 ,\eta )
		=\ave{\left(\frac{{\vec r}(\sigma_1 ,\eta )-{\vec r}(\sigma_2 ,\eta )}{\chi_{\rm S}}\right)^2}_{\rm seg}
		=\frac{1}{\chi_{\rm S}^2}\int^{\sigma_1}_{\sigma_2}\dd\sigma_3\dd\sigma_4\,
			T_{\rm S}(\sigma_3 -\sigma_4 ,\eta )
	\,,\\
	&\Pi_{\rm S}(\sigma_1 -\sigma_2 ,\eta )
		=\ave{\left(\frac{{\vec r}(\sigma_1 ,\eta )-{\vec r}(\sigma_2 ,\eta )}{\chi_{\rm S}}\right)\cdot\dot{\vec r}(\sigma_2 ,\eta )}_{\rm seg}
		=\frac{1}{\chi_{\rm S}}\int^{\sigma_1}_{\sigma_2}\dd\sigma_3\,M_{\rm S}(\sigma_3 ,\eta )
	\,.
}
%-----------------------------------------------------------------------%
It follows that the auto- and cross-power spectra \eqref{eq:Cl_Theta_STheta_S 2} and \eqref{eq:Cl_Theta_Sphi_S 2}
can be recast as
%-----------------------------------------------------------------------%
\al{
	&\ell^2\,C_\ell^{\Theta_{\rm S}\Theta_{\rm S}}
		\approx\frac{(8\pi G\mu )^2}{3\mcA}\,
			\int^{\cCMB}_0\dd\chi\,\frac{\dd V}{\dd\chi}\,n_{\rm S}\,
			\frac{1}{\chi^2}
	\notag\\
	&\quad\quad\quad\quad\quad\times
			\int\dd\sigma_{12}\,
				V_{\rm S}(\sigma_{12},\eta_0 -\chi)\,
			\exp\biggl[-\frac{1}{6}\ell^2\Gamma_{\rm S}(\sigma_{12},\eta_0-\chi)\biggr]\,
			\int\dd\sigma_+
	\,,\label{eq:C_l^TsTs 2}\\
	&\ell^3\,C_\ell^{\Theta_{\rm S}\phi_{\rm S}}
		\approx\frac{(8\pi G\mu )^2\,\ell}{3\mcA}\,
			\int^{\cCMB}_0\dd\chi\,\frac{\dd V}{\dd\chi}\,n_{\rm S}\,
			\frac{\cCMB-\chi}{\cCMB\,\chi^2}
	\notag\\
	&\quad\quad\quad\quad\quad\times
			\int\dd\sigma_{12}\,\Pi_{\rm S}(\sigma_{12},\eta_0-\chi)\,
			\exp\biggl[-\frac{1}{6}\ell^2\Gamma_{\rm S}(\sigma_{12},\eta_0-\chi)\biggr]\,
			\int\dd\sigma_+
	\,,\label{eq:C_l^Tsphis 2}
}
%-----------------------------------------------------------------------%
where we have neglected the $\mcO(c_0^2)$ terms and we have introduced $\sigma_+\equiv (\sigma_1+\sigma_2)/2$\,, $\sigma_{12}\equiv\sigma_1-\sigma_2$\,.
It is useful to introduce the angular scale corresponding to the correlation length of a string segment at $\chi_\mathrm S$\,: $\ell_{\rm co}(\chi_{\rm S})\equiv a(\eta_0-\chi_{\rm S})\,\chi_{\rm S}/\xi(\chi_{\rm S})$\,.
Since the integral $\int\dd\sigma_+$ corresponds to the length of a string segment and the correlators are damped at the large angle $|\sigma_{12}|\gg\xi/a$\,, we can take the regions of integration as $\int\dd\sigma_+/\chi_{\rm S}\approx [2\sqrt{1-v_{\rm rms}^2}\ell_{\rm co}]^{-1}$ and $|\sigma_{12}|/\chi_{\rm S}\leq [2\sqrt{1-v_{\rm rms}^2}\ell_{\rm co}]^{-1}$\,.
We then obtain the angular power spectra as
%-----------------------------------------------------------------------%
\al{
	&\ell^2\,C_\ell^{\Theta_{\rm S}\Theta_{\rm S}}
		\approx \frac{(8\pi G\mu)^2\,v_{\rm rms}^2}{6\mcA\,(1-v_{\rm rms}^2)\,\ell}\,
			\int^{\chi_{\rm CMB}}_0\dd\chi\,\frac{\dd V}{\dd\chi}\,n_{\rm S}
			\frac{1}{\ell_{\rm co}}\,
				U_0\left(\frac{\ell}{2\ell_{\rm co}}\right)
	\,,\label{eq:C_l^TsTs}\\
	&\ell^3\,C_\ell^{\Theta_{\rm S}\phi_{\rm S}}
		\approx \frac{(8\pi G\mu)^2\,c_0}{{12}\mcA\,(1-v_{\rm rms}^2)^2\,\ell^2}\,
			\int^{\cCMB}_0\dd\chi\,\frac{\dd V}{\dd\chi}\,n_{\rm S}\,
			\frac{\cCMB-\chi}{\cCMB}\,
			U_2\left(\frac{\ell}{2\ell_{\rm co}}\right)
	\label{eq:C_l^Tsphis}
}
%-----------------------------------------------------------------------%
with $U_n(s)\equiv\int^s_{-s}\dd t\,t^n\exp (-t^2/6)$\,.
Once the parameters $G \mu$ and $P$ and the scaling values of the string network are given, we can calculate the angular power spectra by {performing the integrations in} eqs.~\eqref{eq:C_l^TsTs} and \eqref{eq:C_l^Tsphis}.
We first evaluate the auto-power spectrum \eqref{eq:C_l^TsTs} for $P=1$ to see the consistency with previous works. 
One can see that its typical amplitude at $\ell =10^3$ is $[\ell (\ell +1)/2\pi ]C_\ell^{\Theta_\rmS\Theta_\rmS}\approx 17(G\mu )^2$\,, 
and it behaves as $\ell^{-1}$ on small scales, while it has a plateau on large scales. 
It is in good agreement with our previous result found with different method in \cite{Yamauchi:2010ms}
and the numerical result by Fraisse \textit{et al.}~\cite{Fraisse:2007nu}.
Hence in the subsequent analysis we use the analytic model to estimate the string correlations.

%>>>>>>>>>>>>>>>>>>>>>>>>>>>>>>>FIGURE<<<<<<<<<<<<<<<<<<<<<<<<<<<<<<<<<<%
\begin{figure}[tbp]
\bc
\begin{tabular}{cc}
\begin{minipage}{0.5\hsize}
\bc
\includegraphics[width=\hsize ]{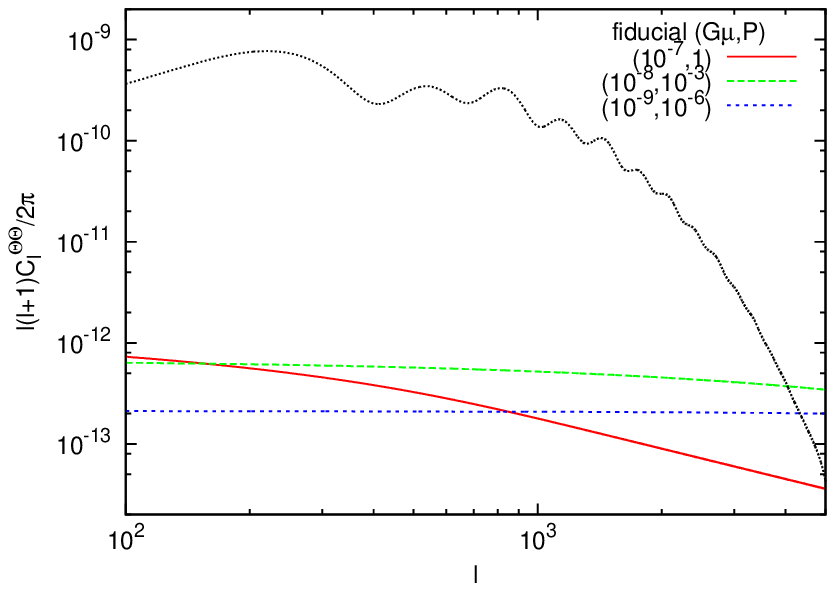}
\ec
\end{minipage}
\begin{minipage}{0.5\hsize}
\bc
\includegraphics[width=\hsize ]{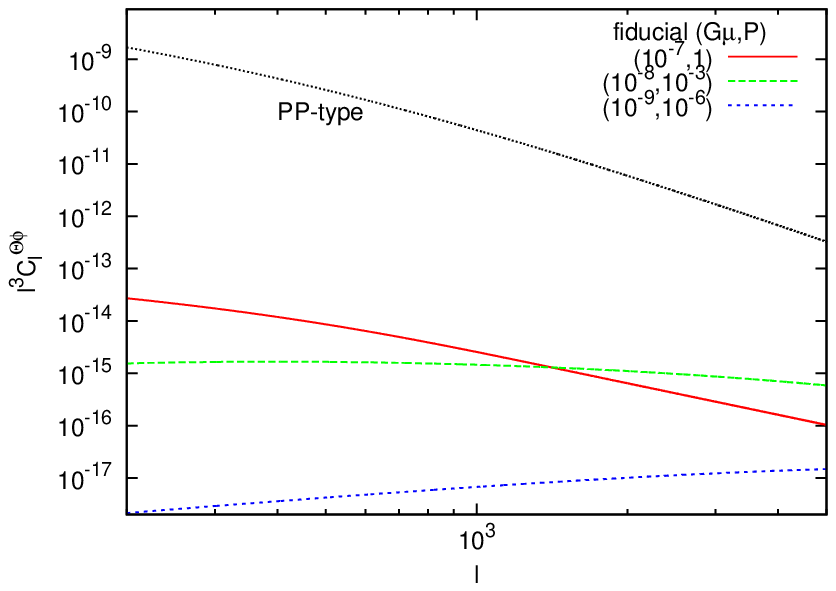}
\ec
\end{minipage}
\end{tabular}
\ec
\caption{
{\it Left}: The auto-power spectra for the temperature fluctuations induced by the Gott-Kaiser-Stebbins (GKS) effect [eq.~\eqref{eq:C_l^TsTs}] with $(G\mu ,P)=(10^{-7},1)$ (red solid)\,, $(10^{-8},10^{-3})$ (green dashed), and $(10^{-9},10^{-6})$ (blue dashed).
For comparison, the spectrum due to the primordial density perturbations is shown in black dotted.
{\it Right}: The cross-correlations between the GKS temperature fluctuations and the string-induced lensing potential [eq.~\eqref{eq:C_l^Tsphis}].
The black dotted line is the ISW-lensing cross correlation due to the primordial density perturbations [eq.~\eqref{eq:C_l^PP}].
}
\label{plot_Pw}
\end{figure}
%>>>>>>>>>>>>>>>>>>>>>>>>>>>>>>>>>><<<<<<<<<<<<<<<<<<<<<<<<<<<<<<<<<<<<<%
In figure~\ref{plot_Pw}, we plot the auto-power spectrum for the GKS temperature fluctuations and the cross correlation between the GKS fluctuations
and the string-induced lensing potential.
To be specific, we consider the three fiducial values of the string parameters: 
$\left(G\mu ,P\right) =\left(10^{-7},1\right)\,,\left( 10^{-8},10^{-3}\right)\,,\left( 10^{-9},10^{-6}\right)$\,.
These fiducial values are still consistent with the recent observation of the small-scale CMB angular power spectrum~\cite{Yamauchi:2010ms}.
Analytic estimation implies that the auto- and cross-power spectra, eqs.~\eqref{eq:C_l^TsTs} and \eqref{eq:C_l^Tsphis}\,, 
roughly scale as $\ell^2\,C_\ell^{\Theta_{\rm S}\Theta_{\rm S}}\propto (G\mu )^2P^{-1/2}\ell^0$\,, 
$\ell^3\,C_\ell^{\Theta_{\rm S}\phi_{\rm S}}\propto (G\mu)^2P^{+1/2}\ell$ 
for $\ell\ll\ell_{\rm co}(\chi_{\rm CMB})\approx 156(\tilde cP/0.23)^{-1/2}$ and
$\ell^2\,C_\ell^{\Theta_{\rm S}\Theta_{\rm S}}\propto (G\mu )^2P^{-1}\ell^{-1}$\,, 
$\ell^3\,C_\ell^{\Theta_{\rm S}\phi_{\rm S}}\propto (G\mu )^2P^{-1}\ell^{-2}$ 
for $\ell\gg\ell_{\rm co}(\chi_{\rm CMB})$\,, respectively.

We will briefly discuss the unlensed angular bispectrum induced by the GKS effect.
From eqs.~\eqref{eq:bispectrum def} and \eqref{eq:GKS effect}\,, the Poisson term of the SSS-type angular bispectrum can be described by
%-----------------------------------------------------------------------%
\al{
	B^{\rm SSS}(\ell_1 ,\ell_2 ,\ell_3 )
		&=\frac{1}{\mcA}\ave{\Theta_{\rm S}(\bm\ell_1)\,\Theta_{\rm S}(\bm\ell_2)\,\Theta_{\rm S}(\bm\ell_3)}
	\notag\\
		&\approx -\frac{i(8\pi G\mu)^3}{\mcA}\,\frac{1}{\ell_1^2\,\ell_2^2\,\ell_3^2}\,
			\int^{\chi_{\rm CMB}}_0\dd\chi\,\frac{\dd V}{\dd\chi}\,
			\left(\prod_a\int\!\dd\psi_a\right)\,f_\mathrm S(\{\psi_a\})
	\notag\\
	&\quad\times
				\int\dd\sigma_1\dd\sigma_2\dd\sigma_3\,
                                \left[
					\prod_{n=1}^3
				(\bm\ell_n\cdot\dot{\bm\theta}_{\rm S}(\sigma_n))
                                \right]\,
					\exp\biggl[\,-i\sum_{m=1}^3
				(\bm\ell_m\cdot\bm\theta_{\rm S}(\sigma_m))
\biggr]
	\,.
}
%-----------------------------------------------------------------------%
Following the same steps as the angular power spectra [\eqref{eq:C_l^TsTs 2} and \eqref{eq:C_l^Tsphis 2}], we can write down the SSS-type bispectrum in terms of the string correlators \eqref{eq:V_S def}-\eqref{eq:M_S def} as
%-----------------------------------------------------------------------%
\al{
	&B^{\rm SSS}(\ell_1 ,\ell_2 ,\ell_3)\notag\\
		&\quad\approx -\frac{(8\pi G\mu)^3}{9\mcA}\,
			\frac{\ell_{12}\,\ell_{31}}{\ell_1^2\,\ell_2^2\,\ell_3^2}\,
			\int^{\cCMB}_0\dd\chi\,\frac{\dd V}{\dd\chi}\,n_{\rm S}\,
			\frac{1}{\chi^3}\,\int\dd\sigma_{123}
	\notag\\
	&\quad\quad\quad\times
			\int\dd\sigma_{12}\dd\sigma_{31}\,
			V_{\rm S}(\sigma_{12},\eta_0-\chi)\,
			\Pi_{\rm S}(\sigma_{31},\eta_0-\chi)
	\notag\\
	&\quad\quad\quad\times
		\exp
			\Biggl[
				-\frac{1}{6}
					\biggl\{
						\ell_{12}\,\Gamma_{\rm S}(\sigma_{12},\eta_0-\chi)
						+\ell_{31}\,\Gamma_{\rm S}(\sigma_{31},\eta_0-\chi)
						+\ell_{23}\,\Gamma_{\rm S}(\sigma_{12}-\sigma_{31},\eta_0-\chi)
					\biggr\}
			\Biggr]
%			F(\ell_{12},\ell_{23},\ell_{31})
	\notag\\
	&\quad\quad
			+\left(\text{perms}\right)
	\,,
}
%-----------------------------------------------------------------------%
where we have neglected the $\mcO(c_0^2)$ terms and we have introduced $\sigma_{mn}\equiv\sigma_m-\sigma_n$\,, $\sigma_{123}\equiv (\sigma_1+\sigma_2+\sigma_3 )/3$\,.
Taking the regions of integration as $|\sigma_{12}|/\chi_{\rm S}<[2\sqrt{1-v_{\rm rms}^2}\,\ell_{\rm co}]^{-1}$\,, $|\sigma_{31}|/\chi_{\rm S}<[2\sqrt{1-v_{\rm rms}^2}\,\ell_{\rm co}]^{-1}$\,, and $\int\dd\sigma_{123}/\chi_{\rm S}\approx [2\sqrt{1-v_{\rm rms}^2}\,\ell_{\rm co}]^{-1}$, and adopting the asymptotic values of the correlators \eqref{eq:asymptotic value of V_S}-\eqref{eq:asymptotic value of Pi_S}, we obtain the approximate form of the SSS-type bispectrum as
%-----------------------------------------------------------------------%
\al{
	B^{\rm SSS}(\ell_1,\ell_2,\ell_3)
		\approx &-\frac{(8\pi G\mu )^3\,v_{\rm rms}^2\,c_0}{36\mcA\,(1-v_{\rm rms}^2)^{5/2}}\,
			\frac{\ell_{12}\,\ell_{31}}{\ell_1^2\,\ell_2^3\,\ell_3^5\,|\sin\theta_{23}|^3}
			\int^{\cCMB}_0\dd\chi\,\frac{\dd V}{\dd\chi}\,n_{\rm S}
	\notag\\
	&\quad\quad\times
			U_0\left(\frac{\ell_2}{2\ell_{\rm co}}\left(1+\frac{|\ell_{23}|}{\ell_2^2}\right)\right)\,
			U_2\left(\frac{\ell_3\,|\sin\theta_{23}|}{2\ell_{\rm co}}\right)
			+\left(\text{perms}\right)
	\label{eq:B^SSS}
}
%-----------------------------------------------------------------------%
with $\cos\theta_{mn}=-\ell_{mn}/\ell_m\ell_n$\,.
To illustrate the typical behavior of the SSS-type bispectrum, we evaluate
the asymptotic form of its equilateral shape, $\ell^4 B^{\rm SSS}(\ell ,\ell ,\ell )$\,, 
and find that it roughly scales as
$(G\mu )^3P\ell^2$ for $\ell\ll\ell_{\rm co}(\chi_{\rm CMB})$ and
$(G\mu )^3P^{-1}\ell^{-2}$ for $\ell\gg\ell_{\rm co}(\chi_{\rm CMB})$\,,
respectively~\footnote{The dependence on $G\mu$ and $\ell$ for large $\ell$ in eq.~\eqref{eq:B^SSS} agrees 
with the results found with somewhat different routes in \cite{Hindmarsh:2009qk,Regan:2009hv,Ringeval:2010ca}.
The SSS-type bispectrum given in this paper cannot explain some features such as the substructure
observed in the more realistic string-induced bispectrum obtained by Planck collaboration~\cite{Ade:2013xla},
which is probably due to the small-scale correlations. 
However, the primary purpose of the present paper is to show the appearance of the string-induced ISW-lensing 
bispectra. In this sense, the construction of a more realistic model of the bispectra is beyond the
scope of the paper. We hope to come back to this issue in a future publication.
}.

%%%%%%%%%%%%%%%%%%%%%%%%%%%%%%%%%%%%%%%%%%%%%%%%%%%%%%%%%%%%%%%%%%%%%%%%%
%=======================================================================%
\subsection{Signal-to-noise ratio}
\label{sec:Signal-to-noise ratio}
%=======================================================================%
%%%%%%%%%%%%%%%%%%%%%%%%%%%%%%%%%%%%%%%%%%%%%%%%%%%%%%%%%%%%%%%%%%%%%%%%%

%>>>>>>>>>>>>>>>>>>>>>>>>>>>>>>>FIGURE<<<<<<<<<<<<<<<<<<<<<<<<<<<<<<<<<<%
\begin{figure}[tbp]
\bc
\includegraphics[width=140mm]{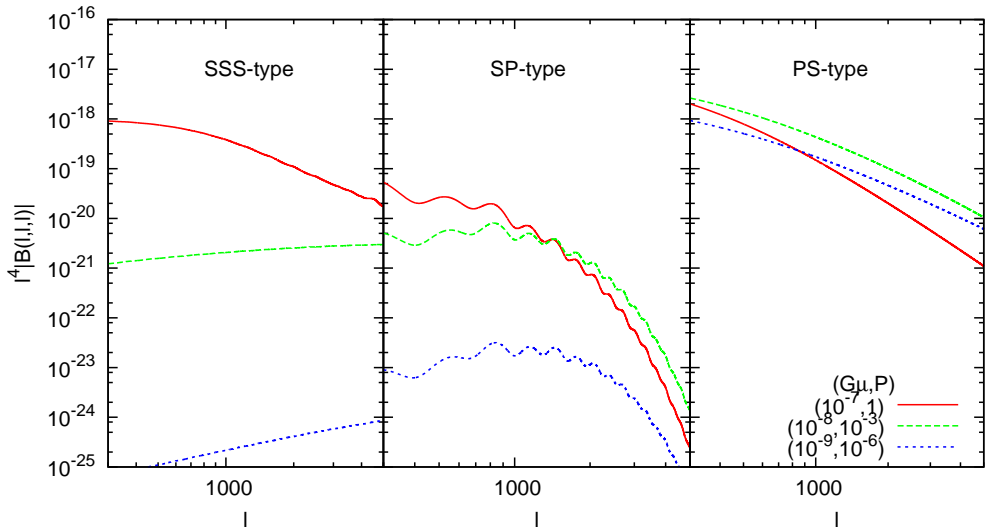}
\caption{
The string-induced bispectra with the equilateral shape, $\ell^4\,B(\ell,\ell,\ell)$\,.
From left to light panels, SSS-type unlensed bispectrum, the SP-type, and PS-type ISW-lensing bispectra.
The curves are for $(G\mu,P)=(10^{-7},1)$ (red solid), $(10^{-8},10^{-3})$ (green dashed), and $(10^{-9},10^{-6})$ (blue dotted).
}
\label{B_eq}
\ec
\end{figure} 
%>>>>>>>>>>>>>>>>>>>>>>>>>>>>>>>>>><<<<<<<<<<<<<<<<<<<<<<<<<<<<<<<<<<<<<%
%>>>>>>>>>>>>>>>>>>>>>>>>>>>>>>>FIGURE<<<<<<<<<<<<<<<<<<<<<<<<<<<<<<<<<<%
\begin{figure}[tbp]
\bc
\includegraphics[width=140mm]{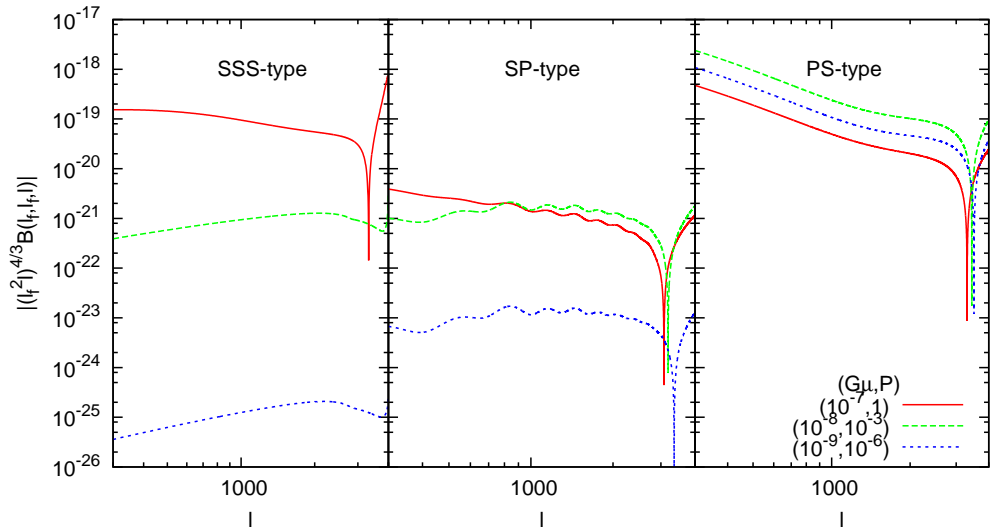}
\caption{
The string-induced bispectra with the isosceles shape, $(\ell_{\rm f}^2\ell )^{4/3}B(\ell_{\rm f},\ell_{\rm f},\ell)$\,, with $\ell_{\rm f}=2000$\,.
From left to light panels, SSS-type unlensed bispectrum, the SP-type, and PS-type ISW-lensing bispectra.
The meaning of the curves and parameters are the same as figure~\ref{B_eq}.
}
\label{B_ang}
\ec
\end{figure} 
%>>>>>>>>>>>>>>>>>>>>>>>>>>>>>>>>>><<<<<<<<<<<<<<<<<<<<<<<<<<<<<<<<<<<<<%
%>>>>>>>>>>>>>>>>>>>>>>>>>>>>>>>FIGURE<<<<<<<<<<<<<<<<<<<<<<<<<<<<<<<<<<%
\begin{figure}[tbp]
\bc
\includegraphics[width=140mm]{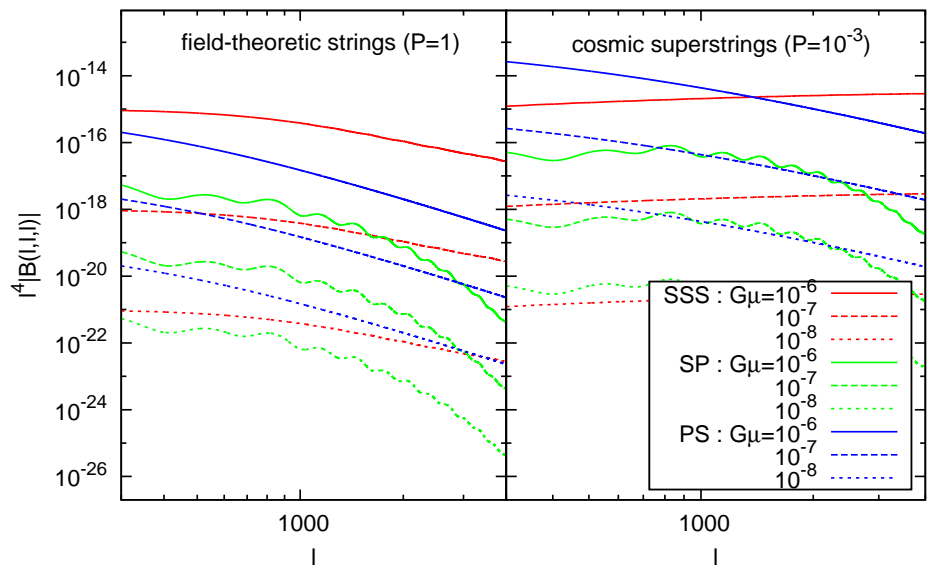}
\caption{
The SSS-type(red), SP-type(green), and PS-type(blue) bispectra for the various values of the string tension.
The curves are for $G\mu =10^{-6}$ (solid), $G\mu =10^{-7}$ (dashed), and $G\mu =10^{-8}$ (dotted).
}
\label{B_compare}
\ec
\end{figure} 
%>>>>>>>>>>>>>>>>>>>>>>>>>>>>>>>>>><<<<<<<<<<<<<<<<<<<<<<<<<<<<<<<<<<<<<%
Based on the formulae derived in the previous subsection, we now discuss the detectability of the CMB signals from a cosmic string network.
Let us first examine the shape of the spectra and the dependence on the string parameters.
Figures~\ref{B_eq} and \ref{B_ang} show the string-induced bispectra with the equilateral and isosceles shapes, namely $\ell^4 B(\ell ,\ell ,\ell )$ and $(\ell_{\rm f}^2\ell)^{4/3}B(\ell_{\rm f},\ell_{\rm f},\ell)$ with $\ell_{\rm f}=2\times 10^3$\,, respectively.
We have specifically set the fiducial values of the string parameters to $\left( G\mu ,P\right) =\left( 10^{-7},1\right)\,, \left(10^{-8},10^{-3}\right)\,, (10^{-9},10^{-6})$\,.
As we mentioned in section~\ref{sec:String correlations}, among the three types of the equilateral bispectra in figure~\ref{B_eq}, the SP-type is particularly suppressed due to the exponential Silk damping, so only the SSS- and PS-type bispectra can be relevant at the small scale.
One can also see that these bispectra are sensitive to the string tension $G\mu$ and the intercommuting probability $P$\,, and 
their dependences on $P$ are rather different.
We then plot the equilateral SSS- and PS-type bispectra as a function of $G\mu$ in figure~\ref{B_compare}\,.
If we consider the ordinary field-theoretic strings ($P=1$) with $G\mu\approx 10^{-7}$\,, the SSS-type bispectrum gives
the dominant contribution. For the strings with the tension smaller than the current upper bound, $G\mu <10^{-7}$, the PS-type 
rather than the SSS-type could dominate the total bispectrum even for $P=1$.
In the case of cosmic superstrings ($P=10^{-3}$),
the situation is not much different from the case of $P=1$, but the cross-over point of the string tension becomes larger.
In particular, as far as we consider the cosmic superstrings with the tension obtained in \cite{Yamauchi:2010ms}, that is $G\mu\lesssim 10^{-8}$ for $P=10^{-3}$\,,
the total string-induced bispectrum is always dominated by the PS-type rather than the SSS-type.

To estimate the feasibility to detect their signals, we quantify the signal-to-noise ratio for the CMB temperature bispectrum.
In the flat-sky approximation, the cumulative signal-to-noise ratio for each CMB bispectrum can be estimated by the optimal inverse-variance weighted statistic as~\cite{Hu:2000ee}
%-----------------------------------------------------------------------%
\al{
	\left(\frac{S}{N}\right)^2_{<\ell_{\rm max}}
		=\frac{1}{4\pi^3}
			\int_{\ell_i\in [\ell_{\rm min},\ell_{\rm max}]}
				\!\!\!\dd^2{\bm\ell}_1\dd^2{\bm\ell}_2\,
				\frac{[B(\ell_1 ,\ell_2 ,\ell_3 )]^2}
				{6(C_{\ell_1}^{\Theta\Theta}+N_{\ell_1}^{\Theta\Theta})\,
					(C_{\ell_2}^{\Theta\Theta}+N_{\ell_2}^{\Theta\Theta})\,
					(C_{\ell_3}^{\Theta\Theta}+N_{\ell_3}^{\Theta\Theta})}
	\,,
}
%-----------------------------------------------------------------------%
where $\ell_3=\sqrt{\ell_1^2+\ell_2^2+2\ell_1\ell_2\cos\theta_{12}}$\,, $N_\ell^{\Theta\Theta}$ is the noise spectrum from the detectors and the residual foreground.
Since we are interested in the flat-sky bispectrum, we have introduced 
the minimum multipole $\ell_{\rm min}$ and we set $\ell_{\rm min}=200$ hereafter.
The instrumental noise is given by
%-----------------------------------------------------------------------%
\al{
	N_\ell^{\Theta\Theta}
		=\biggl[\sum_{\nu}\left( N_{\ell ,\nu}^{\Theta\Theta}\right)^{-1}\biggr]^{-1}
	\quad\text{with}\quad
	N_{\ell ,\nu}^{\Theta\Theta}
		=\left(\frac{\sigma_{\nu ,T}\,\theta_\nu}{T_{\rm CMB}}\right)^2
			\exp\Biggl[
				\frac{\ell\,(\ell +1)\,\theta_\nu^2}{8\ln 2}
			\Biggr]
	\,,
}
%-----------------------------------------------------------------------%
where $T_{\rm CMB}=2.7\,{\rm K}$ is the mean temperature of the CMB, $\theta_\nu$ and $\sigma_{\nu, T}$ represent the beam size and the sensitivity of each channel, respectively.
We summarize the basic parameters for Planck~\cite{:2006uk} and ACTPol~\cite{arXiv:1006.5049} in table~\ref{CMB survey design}.
%TTTTTTTTTTTTTTTTTTTTTTTTTTTTTTTTTTTTTTTTTTTTTTTTTTTTTTTTTTTTTTTTTTTTTTTTT%
\begin{table}[tbp]
\caption{
The experimental specifications for the Planck and ACTPol used in this paper.
The quantity $\theta_\nu$ is the beam size, and $\sigma_{\nu,T}$ represents the sensitivity of each channel to the temperature.
The quantity $\nu$ means a channel frequency.}
\label{CMB survey design}
\bc
\begin{tabular}{ccccc} \hline 
Experiment & $f\rom{sky}$ & $\nu$ [GHz] & $\theta_{\nu}$ [arcmin] 
& $\sigma_{\nu,T}$ [$\mu$K/pixel]
\\ \hline 
Planck~\cite{:2006uk} & 0.65 & 30 & 33 & 4.4 \\ 
 & & 44 & 23 & 6.5 \\ 
 & & 70 & 14 & 9.8 \\
 & & 100 & 9.5 & 6.8 \\
 & & 143 & 7.1 & 6.0 \\
 & & 217 & 5.0 & 13.1 \\
 & & 353 & 5.0 & 40.1 \\ \hline 
ACTPol~\cite{arXiv:1006.5049} & 0.1 & 148 & 1.4 & 3.6 \\ \hline 
\end{tabular}
\ec
\end{table} 
%TTTTTTTTTTTTTTTTTTTTTTTTTTTTTTTTTTTTTTTTTTTTTTTTTTTTTTTTTTTTTTTTTTTTTTTTT%
The noise spectrum for the combination of the large- and small-scale experiments is assumed to have the form (see \cite{Namikawa:2011cs}):
%-----------------------------------------------------------------------%
\al{
	N_{\ell ,{\rm Planck+ACTPol}}^{\Theta\Theta}
		=\left(
			\frac{f_{\rm sky}^{\rm ACTPol}}{\left( N_{\ell ,{\rm ACTPol}}^{\Theta\Theta}\right)^2}
			+\frac{f_{\rm sky}^{\rm Planck}-f_{\rm sky}^{\rm ACTPol}}{\left( N_{\ell ,{\rm Planck}}^{\Theta\Theta}\right)^2}
		\right)^{-1/2}
	\,,
}
%-----------------------------------------------------------------------%
where $f_{\rm sky}^{\rm ACTPol}$ and $f_{\rm sky}^{\rm Planck}$ are the fractional sky coverages of ACTPol and Planck, respectively.

%>>>>>>>>>>>>>>>>>>>>>>>>>>>>>>>FIGURE<<<<<<<<<<<<<<<<<<<<<<<<<<<<<<<<<<%
\begin{figure}[tbp]
\bc
\includegraphics[width=140mm]{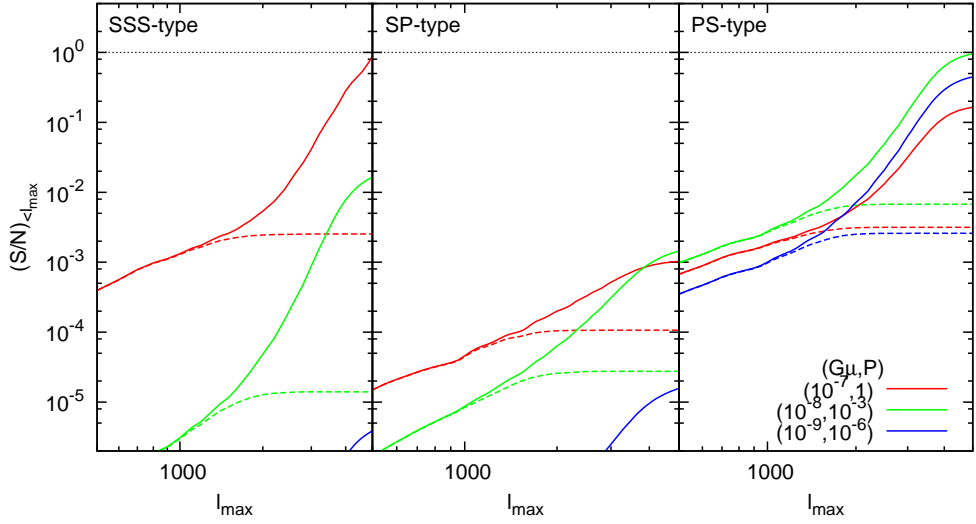}
\caption{
The cumulative signal-to-noise ratios for the bispectra as functions of maximum $\ell$ for the Planck-like experiment (dotted) and the combination of Planck-like and ACTPol-like experiments (solid), respectively.
The curves are for $(G\mu ,P)=(10^{-7},1)$ (red), $(10^{-8},10^{-3})$ (green),
$(10^{-9},10^{-6})$ (blue).
}
\label{bispectrum_SN}
\ec
\end{figure} 
%>>>>>>>>>>>>>>>>>>>>>>>>>>>>>>>>>><<<<<<<<<<<<<<<<<<<<<<<<<<<<<<<<<<<<<%
%>>>>>>>>>>>>>>>>>>>>>>>>>>>>>>>FIGURE<<<<<<<<<<<<<<<<<<<<<<<<<<<<<<<<<<%
\begin{figure}[tbp]
\bc
\includegraphics[width=140mm]{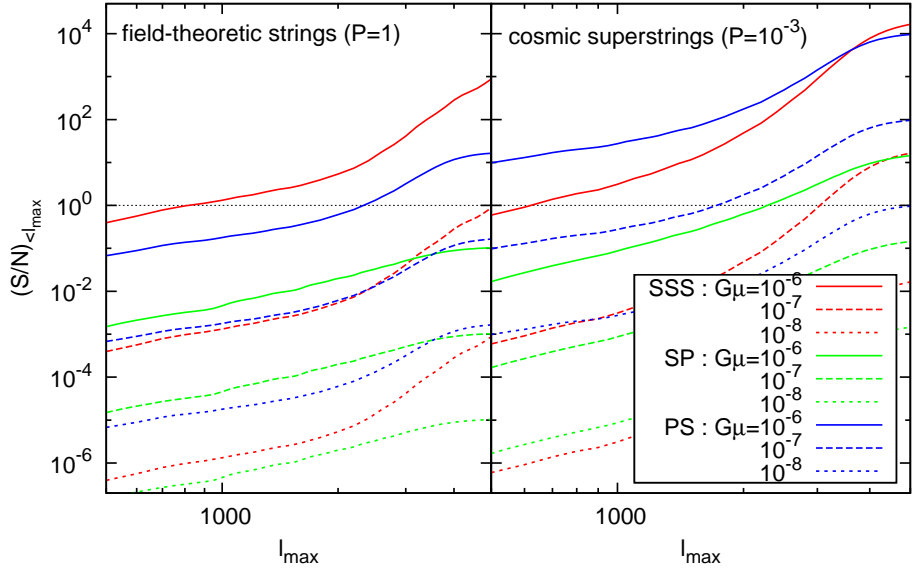}
\caption{
The cumulative signal-to-noise ratios for the bispectra as functions of maximum $\ell$ 
for the various values of the string tension.
The curves are for $G\mu =10^{-6}$ (solid), $G\mu =10^{-7}$ (dashed), and $G\mu =10^{-8}$ (dotted).
}
\label{bispectrum_SN_compare}
\ec
\end{figure} 
%>>>>>>>>>>>>>>>>>>>>>>>>>>>>>>>>>><<<<<<<<<<<<<<<<<<<<<<<<<<<<<<<<<<<<<%
The results for $(S/N)_{<\ell_{\rm max}}$ are shown in figures \ref{bispectrum_SN} and \ref{bispectrum_SN_compare}.
As is expected from figure~\ref{B_eq}, the SP-type bispectrum does not give a significant contribution to the total bispectrum because of the Silk damping, while the SSS- and PS-type bispectra are not damped significantly at small scales and could give the dominant contributions.
The resultant signal-to-noise ratios, namely the detectability of the cosmic strings, are sensitive to the string tension $G\mu$ and intercommuting probability $P$.

%>>>>>>>>>>>>>>>>>>>>>>>>>>>>>>>FIGURE<<<<<<<<<<<<<<<<<<<<<<<<<<<<<<<<<<%
\begin{figure}[tbp]
\bc
\includegraphics[width=100mm]{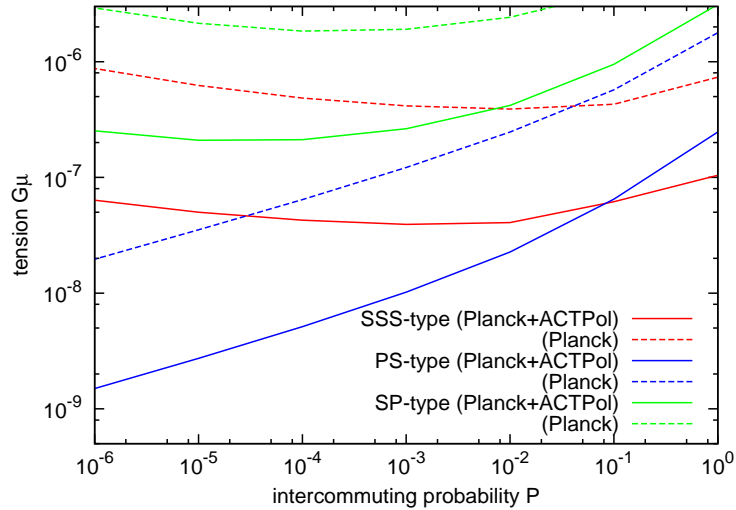}
\caption{
The contours for the signal-to-noise ratio $(S/N)_{<5000}=1$ as a function of the tension $G\mu$ and the intercommuting probability $P$ for the SSS-type (red), the PS-type (green), and the SP-type (blue)\,, respectively.
In each case, $(S/N)_{<5000}$ exceeds $1$ in the region above the contour.
}
\label{bispectrum_Gmu-P_constraints}
\ec
\end{figure} 
%>>>>>>>>>>>>>>>>>>>>>>>>>>>>>>>>>><<<<<<<<<<<<<<<<<<<<<<<<<<<<<<<<<<<<<%
Figure~\ref{bispectrum_Gmu-P_constraints} shows the contour for $(S/N)_{<5000}=1$ as a function of the string tension $G\mu$ and the intercommuting probability $P$\,, where we set the maximum multipole $\ell_{\rm max}$ to $5000$.
An observationally important feature of the string-induced bispectra is that a tighter constraint on $G\mu$ for small $P$ is obtained by using the PS-type than the SSS-type and other ISW-lensing bispectra.
Actually the constraint on $G\mu$ from the SSS-type becomes weaker as $P$ decreases.
This is understood as follows: The SSS-type bispectrum roughly scales as $\propto (G\mu )^3P$ on large scales 
and $(G\mu)^3P^{-1}$ on small scales, as we mentioned in the previous subsection. 
As $P$ decreases, the transition multipole $\ell_{\rm co}(\chi_{\rm CMB})\approx 156\,(\tilde c\,P/0.23)^{-1/2}$ shifts to smaller scale. 
When we choose such small $P$'s that the condition $\ell_{\rm co}(\chi_{\rm CMB})\gg 5000$ is realized, the constraint on $G\mu$ should be determined by the large angle power low, namely $B^{\rm SSS}\propto (G\mu )^3P$\,.
Therefore, the constraint on $G\mu$ from the SSS-type gets weaker for smaller $P$\,.
A similar dependence of the SP-type on $P$ can be also observed.

For $P=1$\,, we could even detect the SSS-type bispectrum with $G\mu\approx 1\times 10^{-7}$ (Planck+ACTPol-like)\,, $7\times 10^{-7}$ (Planck-like)\,, while the signal of the PS-type is detectable for $G\mu\approx 2\times 10^{-7}$ (Planck+ACTPol-like)\,, $2\times 10^{-6}$ (Planck-like)\,.
Hence the predicted constraint on the string tension with $P=1$ from the SSS-type and PS-type bispectra are still consistent with the Planck
measurement~\cite{Ade:2013xla} estimated by using the numerical simulations of Nambu-Goto string network with $P=1$\,.
Furthermore, figure~\ref{bispectrum_Gmu-P_constraints} also implies that when we take account of small-scale observations such as ACTPol, 
the constraint by the string-induced bispectrum would be comparable to those by the power spectrum.
For the smallest intercommuting probability theoretically inferred, $P\approx 10^{-3}$\,, the signal of the PS-type bispectrum 
will be detectable for a string tension $G\mu\approx 1\times 10^{-8}$ (Planck+ACTPol-like)\,, $1\times 10^{-7}$ (Planck-like).

We note that in \cite{Yamauchi:2010ms} we have already obtained the constraint on the string tension for $P\ll 1$ by using
CMB small-scale temperature power spectrum, and in particular the upper bound for $P=10^{-3}$ is given as $G\mu\approx 10^{-8}$\,.
Hence even for the cosmic superstrings ($P\ll 1$), the constraint on $G\mu$ by the PS-type bispectrum
can be competitive with that from the small-scale temperature power spectrum 
in ongoing measurements, such as ACT~\cite{Sievers:2013ica} and SPT~\cite{Dvorkin:2011aj}.

We should emphasize that the models and assumptions given here would be simplistic for a precision study of the CMB observations.
In order to compute the bispectra analytically, we have assumed several idealizations. For instance, we neglected the effect of 
the string motion along the line-of-sight and the higher-order correlation functions; we adopted the very simple models as the evolution
of the string network and the correlations within the string segment; the recombination contributions are dropped.

Although the model and assumption given in this paper might not be realistic enough for an actual string network and further considerations might be needed, the ISW-lensing bispectrum induced by cosmic strings is found to have a new window to constrain the string parameters, $G\mu$ and $P$\,, even more tightly than the GKS-induced bispectrum in the ongoing and future CMB observations.

%%%%%%%%%%%%%%%%%%%%%%%%%%%%%%%%%%%%%%%%%%%%%%%%%%%%%%%%%%%%%%%%%%%%%%%%%
%=======================================================================%
\section{Summary}
\label{sec:Summary}
%=======================================================================%
%%%%%%%%%%%%%%%%%%%%%%%%%%%%%%%%%%%%%%%%%%%%%%%%%%%%%%%%%%%%%%%%%%%%%%%%%

In this paper, we have discussed the effects of the weak gravitational lensing by cosmic strings on the CMB temperature bispectrum.
Additional gravitational sources between the last scattering surface and present can contribute to both the ISW temperature fluctuations and the deflection angle.
The presence of the cross correlation between the ISW temperature fluctuations 
and the lensing potential in general leads to the non-vanishing bispectrum, namely the ISW-lensing bispectrum.
Developing the analytic method to calculate the small-angle correlations for string segments, we can evaluate the auto-angular power spectrum for the ISW temperature fluctuations induced by cosmic strings, namely through the GKS effect, and the cross correlation between the GKS fluctuations and the string-induced lensing potential [eqs.~\eqref{eq:C_l^TsTs} and \eqref{eq:C_l^Tsphis}].

Based on the formulae derived in this paper, we explicitly wrote down the expressions for the string-induced ISW-lensing bispectra (SP- and PS-types) and the GKS-induced bispectrum (SSS-type) [eq.~\eqref{eq:B^SSS}], and estimated the expected cumulative signal-to-noise ratios using the parameters for Planck and ACTPol.
We found that the SSS- and PS-type are dominantly relevant at small scale because the standard ISW-lensing bispectrum (PP-type) and the SP-type bispectrum are exponentially suppressed due to Silk damping.
Thanks to the stronger dependence of the PS-type bispectrum on $P$ than the SSS-type, the PS-type ISW-lensing bispectrum has a new window to constrain the string parameters $G\mu$ and $P$ even more tightly than the SSS-type bispectrum.

The model and prescriptions we employed in this paper may be further improved.
For example, the effect of the string motion along the line of sight was ignored;
the contributions of the connected part of the four-point and higher-order correlation functions were dropped.
However, we would like to emphasize that the generic features are expected to remain the same although those improvements may affect the details of our calculations.

%%%%%%%%%%%%%%%%%%%%%%%%%%%%%%%%%%%%%%%%%%%%%%%%%%%%%%%%%%%%%%%%%%%%%%%%%
%--------------------------- BACK MATTER -------------------------------%
%%%%%%%%%%%%%%%%%%%%%%%%%%%%%%%%%%%%%%%%%%%%%%%%%%%%%%%%%%%%%%%%%%%%%%%%%

%************************* Acknowledgments *****************************%
\acknowledgments

We thank C.-M.~Yoo, T.~Hiramatsu, M.~Yamaguchi, T.~Suyama, S.~Yokoyama, K.~Saikawa, and M.~Hindmarsh 
for valuable comments and useful suggestions.
D.Y.\ is supported by Grant-in-Aid for JSPS Fellows No.~259800.
K.T.\ is supported by JSPS Grant-in-Aid for Young Scientists (B) No.~23740179,
by MEXT through Grant-in-Aid for Scientific Research on Innovative Areas No.~24111710,
and partially by JSPS Grant-in-Aid for Scientific Research (B) No.~24340048.
Y.S.\ is supported in part by MEXT through Grant-in-Aid Scientific Research
on Innovative Areas No.~24111701.

%***************************** appendix ********************************%

%**************************** References *******************************%
%%%%%%%%%%%%%%%%%%%%%%%%%%%%%%%%%%%%%%%%%%%%%%%%%%%%%%%%%%%%%%%%%%%%%%%%%%%%%%%

\end{document}